%% file: main.tex
\documentclass{ieeeaccess}
\usepackage{cite}
\usepackage{amsmath,amssymb,amsfonts}
\usepackage{algorithmic}
\usepackage{graphicx}
\usepackage{textcomp}
\def\BibTeX{{\rm B\kern-.05em{\sc i\kern-.025em b}\kern-.08em
    T\kern-.1667em\lower.7ex\hbox{E}\kern-.125emX}}

\input{preamble}

\input{paper-specific}

\begin{document}
\history{Date of publication xxxx 00, 0000, date of current version xxxx 00, 0000.}
\doi{10.1109/TQE.2023.DOI}
\title{FPGA-based Distributed Union-Find Decoder for Surface Codes}
\author{\uppercase{Namitha Liyanage\authorrefmark{1},
Yue Wu\authorrefmark{1}, Siona Tagare\authorrefmark{1}, and Lin Zhong}.\authorrefmark{1}}
\address[1]{Department of Computer Science, Yale University, New Haven, CT, USA (email: first.last@yale.edu)}
\tfootnote{This work was supported in part by Yale University and NSF MRI Award \#2216030.}

\markboth
{Liyanage \headeretal: FPGA-based Distributed Union-Find Decoder for Surface Codes}
{Liyanage \headeretal: FPGA-based Distributed Union-Find Decoder for Surface Codes}

\corresp{Corresponding author: Namitha Liyanage (email: namitha.liyanage@yale.edu).}

\begin{abstract}
A fault-tolerant quantum computer must decode and correct errors faster than they appear to prevent exponential slowdown due to error correction. The Union-Find (UF) decoder is promising with an average time complexity slightly higher than $O(d^3)$. We report a distributed version of the UF decoder that exploits parallel computing resources for further speedup. Using an FPGA-based implementation, we empirically show that this distributed UF decoder has a sublinear average time complexity with regard to $d$,  given $O(d^3)$ parallel computing resources. The decoding time per measurement round decreases as $d$ increases, the first time for a quantum error decoder. The implementation employs a scalable architecture called Helios that organizes parallel computing resources into a hybrid tree-grid structure. Using a Xilinx VCU129 FPGA, we successfully implement $d$ up to 21 with an average decoding time of 11.5 ns per measurement round under 0.1\% phenomenological noise, and 23.7 ns for $d=17$ under equivalent circuit-level noise. This performance is significantly faster than any existing decoder implementation. Furthermore, we show that \name can optimize for resource efficiency by decoding $d=51$ on a Xilinx VCU129 FPGA with an average latency of 544ns per measurement round.


\end{abstract}


\titlepgskip=-15pt

\maketitle

\input{introduction}
\input{background}
\input{design}

\input{system}

\input{implementation}

\input{results}
\input{related_work}

\input{conclusion}

\appendices

\input{fpga_algo}

\bibliographystyle{IEEEtran}
\bibliography{abr-long,ref}

\EOD

\end{document}

%% file: preamble.tex
\usepackage{setspace}
\usepackage{hyperref}
\usepackage{amsmath, amsthm}

\usepackage[normalem]{ulem}

\usepackage{graphicx}
\usepackage{comment}

\usepackage{listings}
\usepackage[font=footnotesize]{caption}
\usepackage[font=footnotesize]{subcaption}
\usepackage{multirow} 

\usepackage{url}

\newcommand{\secref}[1]{\S\ref{#1}}

\newcommand{\lineref}[1]{L\ref{#1}}

\usepackage{enumitem}

%% file: paper-specific.tex
\usepackage{soul}

\usepackage[ruled,linesnumbered]{algorithm3e}
\renewcommand{\algorithmautorefname}{Algorithm}
\usepackage{balance}

\newcommand{\code}[1]{\texttt{#1}}

\newcommand{\remove}[1]{}

\definecolor{lightblue}{rgb}{0.9,0.9,1}
\definecolor{aqua}{rgb}{0.0, 1.0, 1.0}

\newcommand\name[0]{Helios\xspace}

%% file: introduction.tex
\section{Introduction}
\label{sec:introduction}

The high error rates of quantum devices pose a significant obstacle to realizing a practical quantum computer.  As a result, developing effective quantum error correction (QEC) mechanisms is crucial for successfully implementing a fault-tolerant quantum computer.

One promising approach for QEC is surface codes~\cite{dennis2002topological, fowler2012surface, bonilla2020xzzx} in which information of a single qubit (called a logical qubit) is redundantly encoded across many physical data qubits, with a set of ancillary qubits interacting with the data qubits.
One can detect and potentially correct errors in physical qubits by periodically measuring the ancillary qubits.

Once errors have been detected by measuring ancillary qubits, a classical algorithm, or \emph{decoder}, guesses the underlying error pattern and corrects it accordingly.
The faster errors can be corrected, the more time a quantum computer can spend on useful work. 
Due to the error rate of state-of-the-art qubits, very large surface codes ($d\approx27$) are necessary to achieve fault-tolerant quantum computing~\cite{fowler2012surface, Chen2021Exponential, Gidney2021rsa}. 
Here, $d$ is the distance of the code and is the minimum number of bit or phase flips of physical qubits needed to change the state of the logical qubit.
See \S\ref{sec:background} for more background. 

As surveyed in \S\ref{sec:related}, previously reported decoders capable of decoding errors as fast as measured, or \emph{backlog-free}, either exploit limited parallelism~\cite{das2021liliput, das2022afs, higgott2023sparse, Vittal2023Astrea, barber2023realtime}, or sacrifice accuracy~\cite{Holmes2020Nisq, Ueno2021Qecool, Liao2023WitGreedy}. 
Sparse Blossom
~\cite{higgott2023sparse} and Fusion Blossom~\cite{wu2023Fusion} feature an important algorithmic breakthrough in realizing MWPM-based decoders. 
Fusion Blossom can additionally leverage measurement round-level parallelism to meet the throughput requirement of very large $d$. 
Due to their software-based implementations, Sparse Blossom and Fusion Blossom suffer from decoding times per round that are orders of magnitude longer than this work, especially at larger $d$ and higher noise levels.
When used in a quantum computer, the computer would spend most of the execution time waiting for error correction results.

In this paper we report a \textit{distributed Union-Find (UF) decoder} (\S\ref{sec:Design}) and its FPGA implementation called \textit{\name} (\S\ref{sec:System}). Given $O(d^3)$ parallel resources, our decoder achieves sublinear average time complexity according to empirical results for $d$ up to 21, the first to the best of our knowledge.
Notably, adding more parallel resources will not reduce the decoder's time complexity due to the inherent nature of error patterns.
Our decoder is a distributed design of and logically equivalent to the UF decoder first proposed in~\cite{delfosse2017almost}. 

We implement the distributed Union-Find decoder using \name, a scalable architecture that efficiently organizes parallel computation units. 
\name also allows for a customizable balance between latency and resource usage, adapting to specific requirements.
\name is the first architecture of its kind that can scale to arbitrarily large surface codes by exploiting parallelism at the vertex level of the model graph.
In \S\ref{sec:results}, we present experimental validations of the distributed UF decoder and \name using a VCU129 FPGA board~\cite{vcu129} for various values of $d$ up to 51.
When optimized for decoding time, the decoder can decode $d$ up to 21 for phenomenological noise and $d$ up to 17 for circuit-level noise.
The
average decoding time per measurement round under 0.1\% noise level is 11.5~ns and 21.3~ns, respectively.
When optimized for resource usage, the decoder can decode $d$ up to 51, with an average decoding time of 543.9 ns per measurement round under phenomenological noise of $p=0.001$ for $d=51$.
These results show that our decoder is significantly faster than any existing decoder implementation.
Our results also successfully demonstrate, for the first time, a decoder design with a decreasing average time per measurement round when $d$ increases.
This shows evidence that the decoder can scale to arbitrarily large surface codes without a growing backlog.
In summary, we report the following contributions in this paper. 
\begin{itemize}
    \item A distributed algorithm that implements the Union-Find decoder that can exploit parallel computing units to stop decoding time per measurement round from growing with the code distance $d$.

    \item The \name architecture and its FPGA-based implementation that realize the distributed Union-Find decoder. 

    \item A set of empirical data based on the FPGA implementation that demonstrates decreasing decoding time per round as $d$ grows up to $d=21$ on a VCU 129 FPGA. 
    
    \item A set of empirical data that demonstrates \name can trade-off resource usage for latency by decoding $d=51$ on VCU129 FPGA.
\end{itemize}

\name is open-source and available from~\cite{qecGithub}.

A prior version of this work appeared in ~\cite{liyanage2023scalable}. This work contains development to the previous work by modifying \name to support circuit-level noise, erasure errors, sliding window decoding, and tradeoff latency for lower resource usage using context-switching.

%% file: background.tex
\section{Background}
\label{sec:background}
\input{background_sc.tex}

\input{background_decoding.tex}

%% file: background_sc.tex
\subsection{Error Correction and Surface Code}
\label{ssec:sc}

Quantum Error Correction (QEC) is more challenging than classical error correction due to the nature of Quantum bits (qubits). 
First, qubits cannot be copied to achieve redundancy due to the no-cloning theorem. 
Second, the values of qubits cannot be directly measured as measurements perturb the state of qubits. 
Therefore, QEC is achieved by encoding the \emph{logical state} of a qubit as a highly entangled state of many physical qubits.
Such an encoded qubit is called a \emph{logical} qubit.

\begin{figure} [!t]
	\centering
 \begin{minipage}{.28\textwidth}
    \centering
	\begin{subfigure}{.98\textwidth}
	    \centering
        \includegraphics[width=1\textwidth]{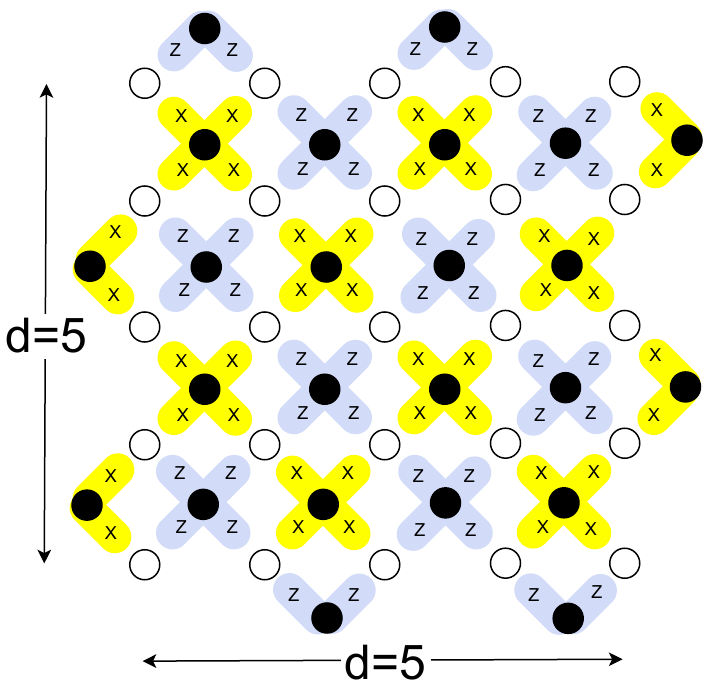}
        \caption{}
        \label{fig:basic_CSS}
    \end{subfigure}
\end{minipage}
 \begin{minipage}{.18\textwidth}
    \centering
	\begin{subfigure}{.98\textwidth}
	    \centering
        \includegraphics[width=1\textwidth]{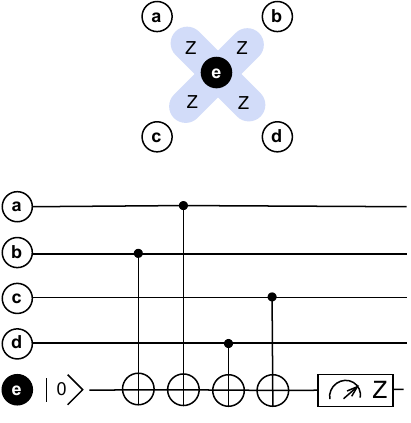}
        \caption{}
        \label{fig:CSS_Z_circuit}
    \end{subfigure}
    \hfill
	\begin{subfigure}{.98\textwidth}
	    \centering
        \includegraphics[width=1\textwidth]{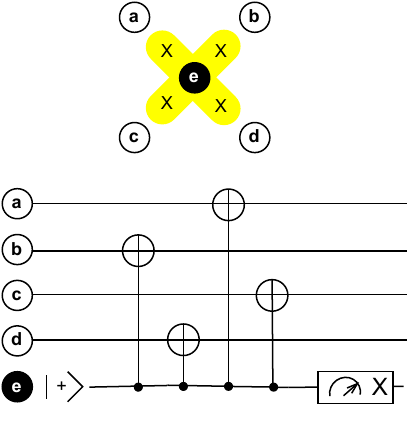}
        \caption{}
        \label{fig:CSS_X_circuit}
    \end{subfigure}
\end{minipage}
	\caption{(a) : Rotated CSS surface code ($d=5$), a commonly used type of surface code. The white circles are data qubits and the black are the Z-type and X-type ancillas. (b) and (c): Measurement circuit of Z-type and X-type ancillas. Excluding the ancillas in the border, each Z-type and X-type ancilla interacts with 4 adjacent data qubits.}
	\label{fig:surfacecode}
\end{figure}

The surface code is the widely used error correction code for quantum computing due to its high error correction capability and ease of implementation due to only requiring connectivity between adjacent qubits. 
A distance $d$ rotated surface code is a topological code made out of $2d^2 - 1$ physical qubits arranged as shown in \autoref{fig:surfacecode}.
A key feature of surface codes is that a larger $d$ can exponentially reduce the rate of logical errors, making them advantageous.
For example, even if the physical error rate is 10 times below the threshold, $d$ should be greater than 17 to achieve a logical error rate below $10^{-10}$~\cite{fowler2012surface}.

A surface code contains two types of qubits, namely data qubits and ancilla qubits.
The data qubits collectively encode the \emph{logical state} of the qubit.
The ancilla qubits (called X-type and Z-type) entangle with the data qubits and by periodically measuring the ancilla qubits, physical errors in all qubits can be potentially discovered and corrected.
An X error occurring in a data qubit will flip the measurement outcome of Z ancilla qubits connected with the data qubit and a Z error will flip the X ancilla qubits likewise.

\textbf{Noise model : }
A noise model defines the types and locations of X and Z errors in a surface code. 
The two prevalent models are the phenomenological noise model and the circuit-level noise model. 
In the phenomenological model, errors are confined to data qubits and occur before gate execution, consequently flipping adjacent ancilla qubits in the same measurement round. 
Conversely, in the circuit-level model, errors can arise between gates, resulting in the flipping of only one adjacent ancilla qubit in the current measurement round and the other in the subsequent round. Additionally, in the circuit-level noise model, errors may occur in ancilla qubits, which effectively equates to errors in the two adjacent data qubits. 
Such errors are termed \textit{hook errors}.

Besides these errors, there are measurement errors and erasure errors. A measurement error occurs when an ancilla qubit produces an erroneous reading due to faults in the readout process. An erasure error represents a detectable leakage of the quantum state in a qubit.

\begin{figure} [!t]
	\centering
\centering
    \begin{minipage}{.30\textwidth}
    \centering
    \begin{subfigure}{0.98\textwidth}
	    \centering
        \includegraphics[width=1\textwidth]{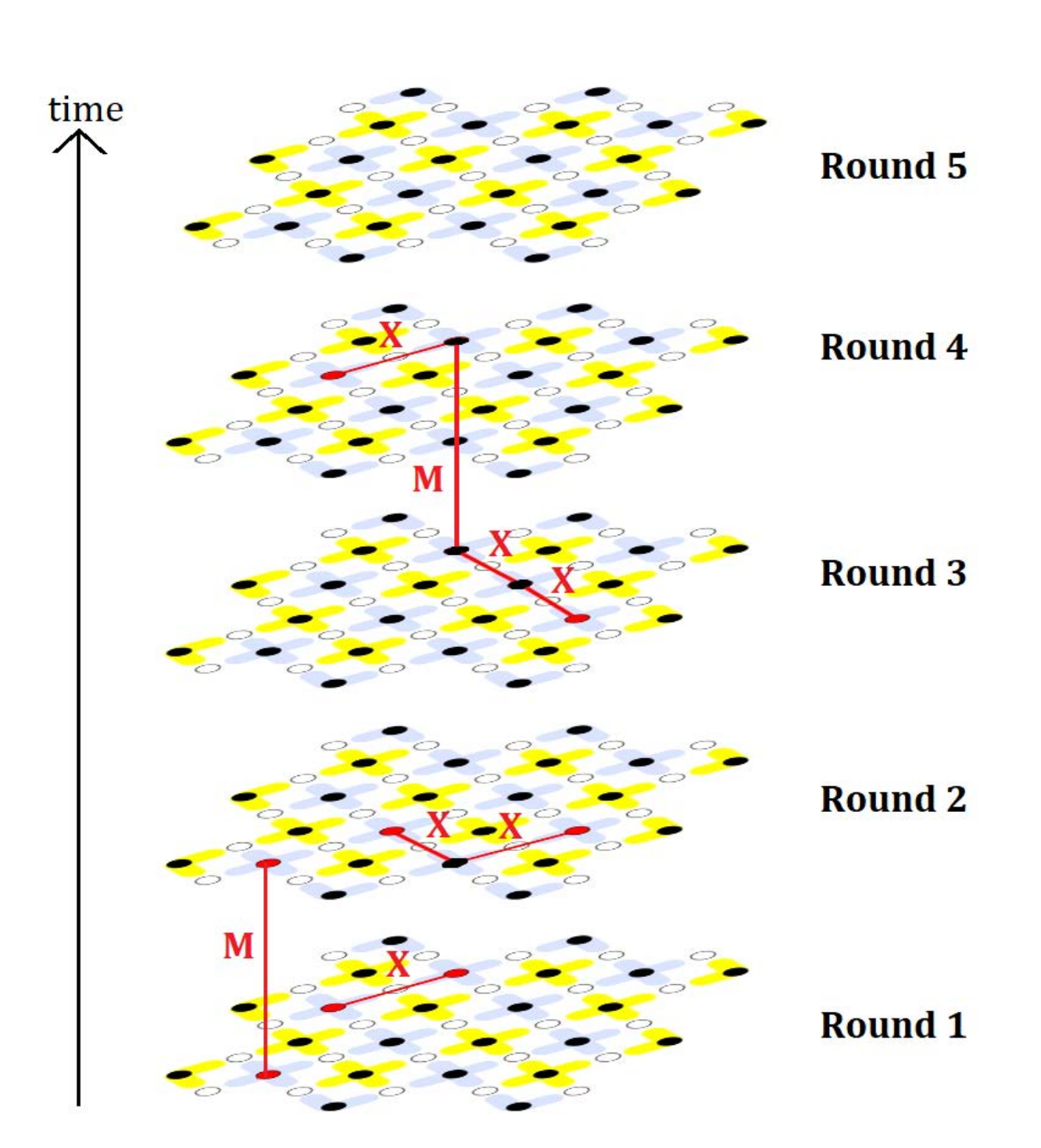}
        \caption{}
        \label{fig:example_syndrome}
        \end{subfigure}
        \end{minipage}
    \begin{minipage}{0.14\textwidth}
    \centering
        	\begin{subfigure}{0.98\textwidth}
	    \centering
        \includegraphics[width=\textwidth]{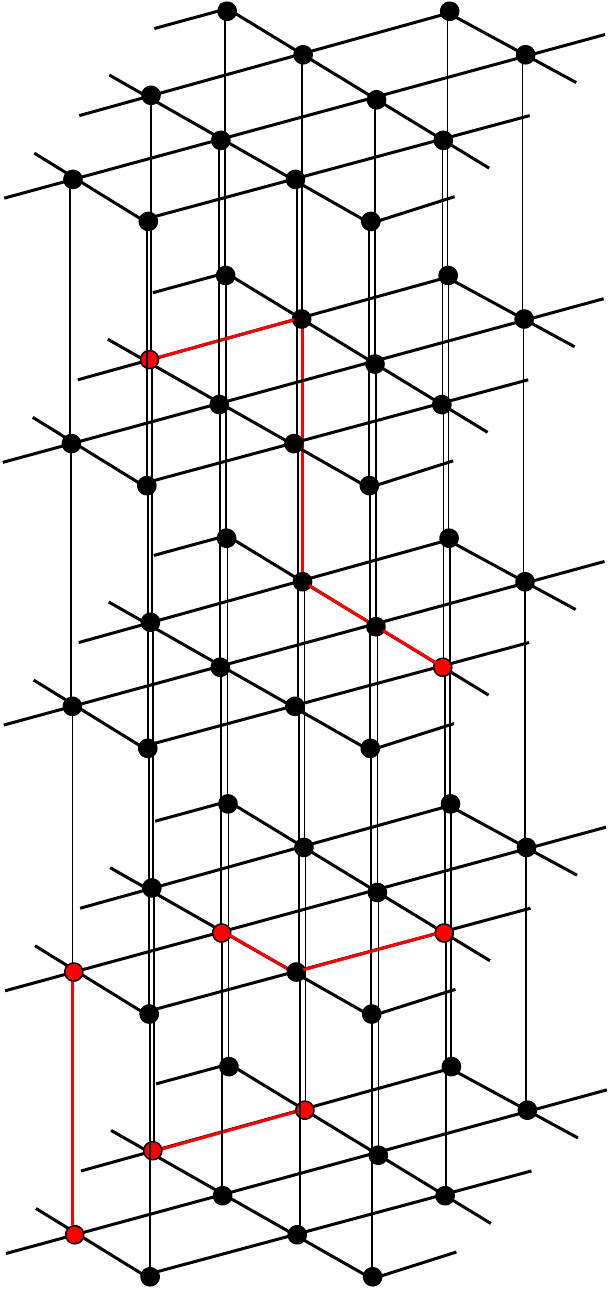}
        \caption{}
        \label{fig:decoding_graph}
    \end{subfigure}
    \end{minipage}
    \begin{minipage}{0.20\textwidth}
    \centering
        	\begin{subfigure}{0.98\textwidth}
	    \centering
        \includegraphics[width=\textwidth]{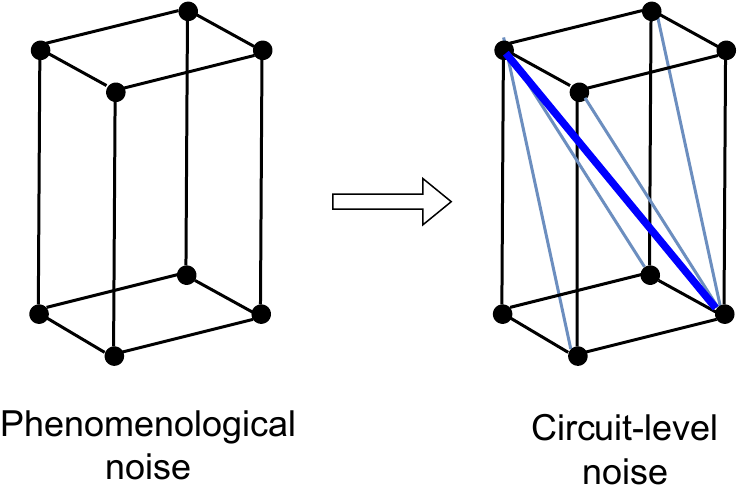}
        \caption{}
        \label{fig:cct_graph_extension}
    \end{subfigure}
    \end{minipage}
	\caption{(a) : An example syndrome of Z stabilizers for $d=5$ surface code with 5 rounds of measurements. The syndrome contains an isolated X-error (round 1), an isolated measurement error (rounds 1 and 2), a chain of two X errors (round 3), and a chain containing X errors and measurement errors spanning multiple measurement rounds (rounds 3 and 4). (b) : Phenomenological noise decoding graph with defect vertices marked red for the syndrome in (a). (c) : Modification of decoding graph from phenomenological noise to circuit level shown only for 8 adjacent vertices. Extra edges in the circuit-level noise decoding graph are shown in blue. The thick blue edge represents a hook error and others represent X-errors spanning two measurement rounds. }
 
	\label{fig:error_patterns}
\end{figure}

A measurement outcome of a flipped ancilla qubit is called a \emph{defect measurement}.
The outcomes from multiple rounds of measurements of ancilla qubits constitute a \emph{syndrome}.
In practice, a syndrome consists of at least $d$ measurement rounds.
\autoref{fig:example_syndrome} shows a syndrome with sample physical qubit errors and shows how they are detected by ancilla qubits.
We only show X errors and measurement errors on Z-type ancillas because Z errors and measurement errors on X-type ancillas can be independently dealt with in the same way.

A syndrome can be conveniently represented by a graph called \emph{decoding graph} in which each vertex represents a measurement outcome of an ancilla.
An edge in this graph corresponds to an independent error source, linking two vertices that represent the defective measurement outcomes caused by this error. 
Thus, the number of edges in the graph depends on the error model under consideration and the weight of an edge is determined by the probability of the error corresponding to the edge.

The decoding graph typically contains $(d+1)\times ((d-1)/2) \times d$ vertices. \autoref{fig:decoding_graph} illustrates the decoding graphs for the phenomenological noise model and \autoref{fig:cct_graph_extension} illustrates how it will be extended for the circuit-level noise model. Notably, in the phenomenological noise model, each vertex has a maximum of 6 incident edges. Conversely, in the circuit-level noise model, a vertex can have up to 12 incident edges.

%% file: background_decoding.tex
\subsection{Error Decoders}
\label{sec:decoder}
Given a syndrome, an error decoder identifies the underlying error pattern,  which will be used to generate a correction pattern. 
As multiple error patterns can generate the same syndrome, the decoder has to make a probabilistic guess of the underlying physical error.
The objective is that when the correction pattern is applied, the chance of the surface code entering a different logical state (i.e a logical error) will be minimized. 

\paragraph{Metrics} The two important aspects of decoders are accuracy and speed.
A decoder must correct errors faster than syndromes are produced to avoid a backlog. 
A faster decoder allows faster execution of a quantum computer, reducing the idle time waiting for decoding to be available.
The average decoding time per measurement round is a widely used criterion for speed.

A decoder must make a careful tradeoff between speed and accuracy.  A faster decoder with lower accuracy requires a larger $d$ to achieve any given logical error rate, which may require more computation overall.

 \paragraph{Union-Find (UF) Decoder} The UF decoder is a fast surface code decoder design first described by Delfosse and Nickerson~\cite{delfosse2017almost}. 
According to \cite{yue2021interpretation}, it can be viewed as an approximation to the blossom algorithm that solves minimum-weight perfect matching (MWPM) problems.
It has a worst-case time complexity of $O(d^3\alpha (d))$, where $\alpha$ is the inverse of Ackermann’s function, a slow-growing function that is less than three for any practical code distances. Based on our analysis, it has an average case time complexity slightly higher than $O(d^3)$.

\autoref{alg:serialuf} describes the UF decoder. It takes a decoding graph $\mathcal{G}(\mathbf{V},\mathbf{E})$ as input.
Each edge $e\in \mathbf{E}$ has a weight and a growth, denoted by $e.w$ and $e.g$, respectively. $e.g$ is initialized with $0$ and the decoder may grow $e.g$ until it reaches $e.w$. When that happens, we say the edge is \emph{fully grown}.

The decoder maintains a set of odd clusters, denoted by $\mathcal{L}$. $\mathcal{L}$ is initialized to include all $\{v\}$ that $v\in\mathbf{V}$ are defect measurements (\lineref{line:initial_list}). 
Each cluster $C$ keeps track of whether its cardinality is odd or even as well as its root element.

The UF decoder iterates over growing and merging the odd cluster list until there are no more odd clusters (inside the \textbf{while} loop of \autoref{alg:serialuf}). 
Each iteration has two stages: Growing and Merging. 
In the \emph{Growing} stage, each odd cluster ``grows'' by increasing the \emph{growth} of the edges incidental to its boundary. 
This process creates a set of \emph{fully grown} edges $\mathcal{F}$ (\lineref{line:grow-start} to \lineref{line:grow-end}). 
The Growing stage is the more time-consuming step as it requires traversing all the edges in the boundary of all the odd clusters and updating the global edge table. 
Since the number of edges is $O(d^3)$, the UF decoder is not scalable for surface codes with large $d$. 

In the \emph{Merging} stage, the decoder goes through each fully-grown edge to merge the two clusters connected by the edge using UNION($u,v$) operation.
The UNION($u,v$) merges the two clusters containing $u$ and $v$ by assigning a 
common root element to the two clusters.
When two clusters merge, the new cluster may become even. 

When there are no more odd clusters, the decoder finds a correction within each cluster and combines them to produce the correction pattern (\lineref{line:peeling}).

\begin{algorithm} [!t]
\DontPrintSemicolon
\caption{Union Find Decoder}\label{alg:serialuf}
\footnotesize
\setcounter{AlgoLine}{0}
\SetKwInOut{Input}{input}
\SetKwInOut{Output}{output}
\Input{A decoding graph $\mathcal{G}(\mathbf{V}, \mathbf{E})$ with X (or Z) syndrome}
\Output{A correction pattern}
\% Initialization\\
    \ForEach{$v\in \mathbf{V}$}{
        \If{$v$ is defect measurement}{
            Create a cluster $\{v\}$
        }
        \label{line:initial_list}
    }

\While{there is an odd cluster}{ \label{line:loop_start}
    \% Growing\\
    $\mathcal{F}\gets \emptyset$\\
    \ForEach{odd cluster $C$}{
    \label{line:grow-start}
         \ForEach{$e=<u,v>$, $u\in C, v\not\in C$ }{
            \If{$e.growth < e.w$}{ 
                $e.growth\gets e.growth+1$ \label{line:grow_serial}\\
                \If{$e.growth = e.w$}{ 
                    $\mathcal{F} \gets \mathcal{F} \cup \{e\}$ \label{line:fusion_set}
                }
            }
        }
        
    }\label{line:grow-end}
    \% Merging\\
    \ForEach{$e = <u,v> \in \mathcal{F}$}{
        UNION($u$, $v$) \label{line:merge}

    }
}
\label{line:loop_end}
\textup{Build correction within each cluster by constructing a spanning tree} \label{line:peeling}
\end{algorithm}

%% file: design.tex
\section{Distributed UF Decoder Design}
\label{sec:Design}

Our goal to build a QEC decoder is scalability to the number of qubits.
As surface codes can exponentially reduce logical error rate with respect to $d$, larger surface codes with hundreds or even thousands of qubits are necessary for fault-tolerant quantum computing. Therefore, the average decoding time per measurement round  should not grow with $d$, to avoid exponential backlog for any larger $d$.

We choose the UF decoder for two reasons.
First, it has a much lower time complexity than the MWPM algorithm. Although in general, the UF decoder achieves lower decoding accuracy than MWPM decoders, it is as accurate in many interesting surface codes and noise models~\cite{yue2021interpretation,huang2020fault}.
Second, the UF decoder maintains fewer intermediate states, which makes it easier to implement in a distributed manner.
We observe that the Growing stage from \lineref{line:grow-start} to \lineref{line:grow-end} in \autoref{alg:serialuf} operates on each vertex independently without dependencies from other vertices. 
A vertex requires only the parity of the cluster it is a part of for the growing stage.
Second, during the merging stage, a vertex only needs to interact with its immediate neighbors (\lineref{line:merge}).

\input{dalgorithm}

\begin{figure*} [!t]
	 \begin{minipage}{0.48\textwidth}   
     \centering
        \includegraphics[width=0.72\linewidth]{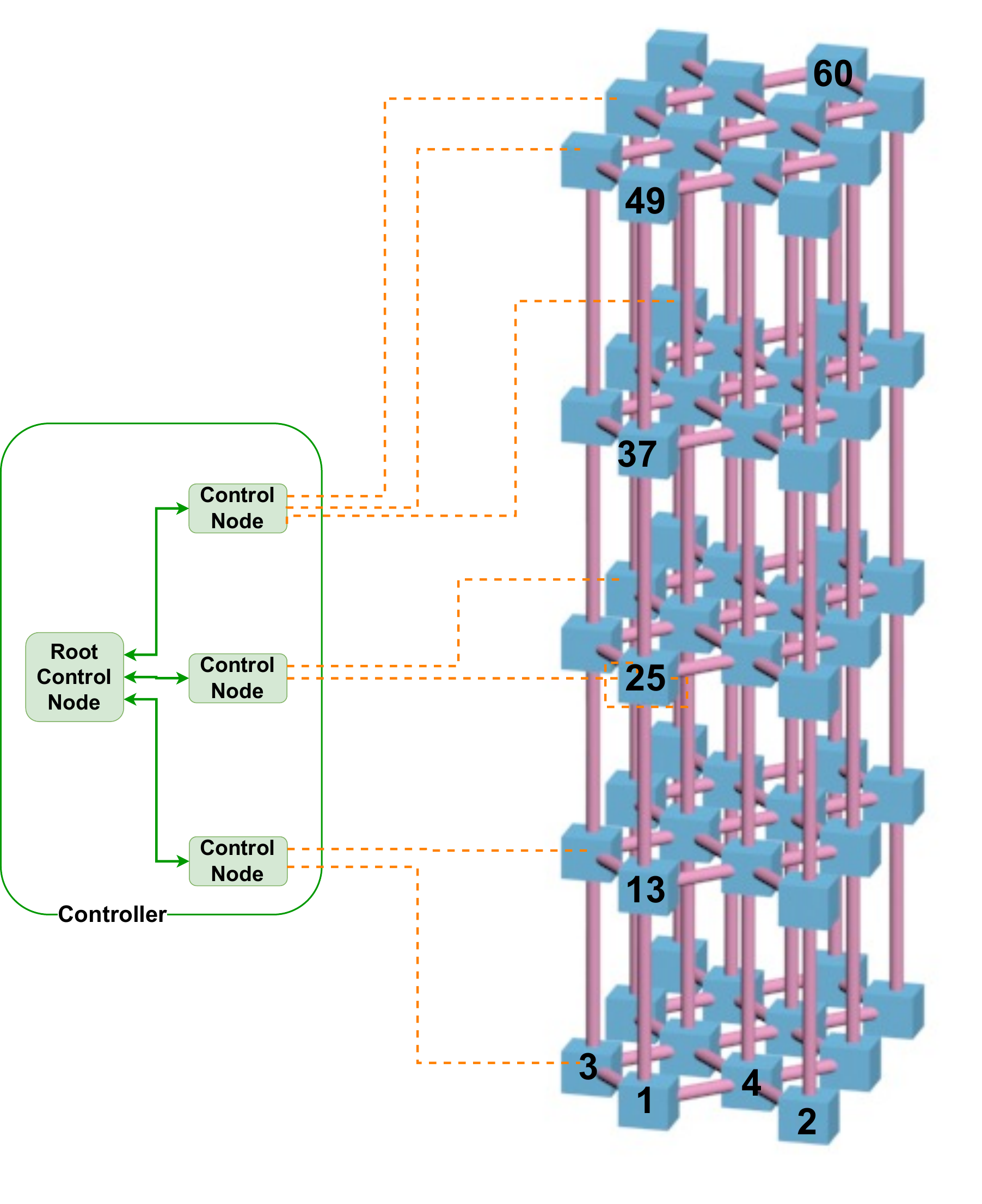}
	\caption{\name architecture for d=5 surface code for 5 measurement rounds for phenomenological noise model. As d=5 surface code has 12 ancilla qubits of Z-type, \name contains a 12x5 PE array. PE $n$ indicates PE with $v.id=n$. Not all links from the controller to PEs and all $v.id$s are shown in the figure. The architecture for circuit-level noise has additional links between PEs corresponding to the additional edges in the decoding graph of circuit-level noise} 
 
 	\label{fig:pe_array}
\end{minipage}
\hfill
\begin{minipage}{0.48\textwidth}
    	    \centering
        \includegraphics[width=0.80\linewidth]{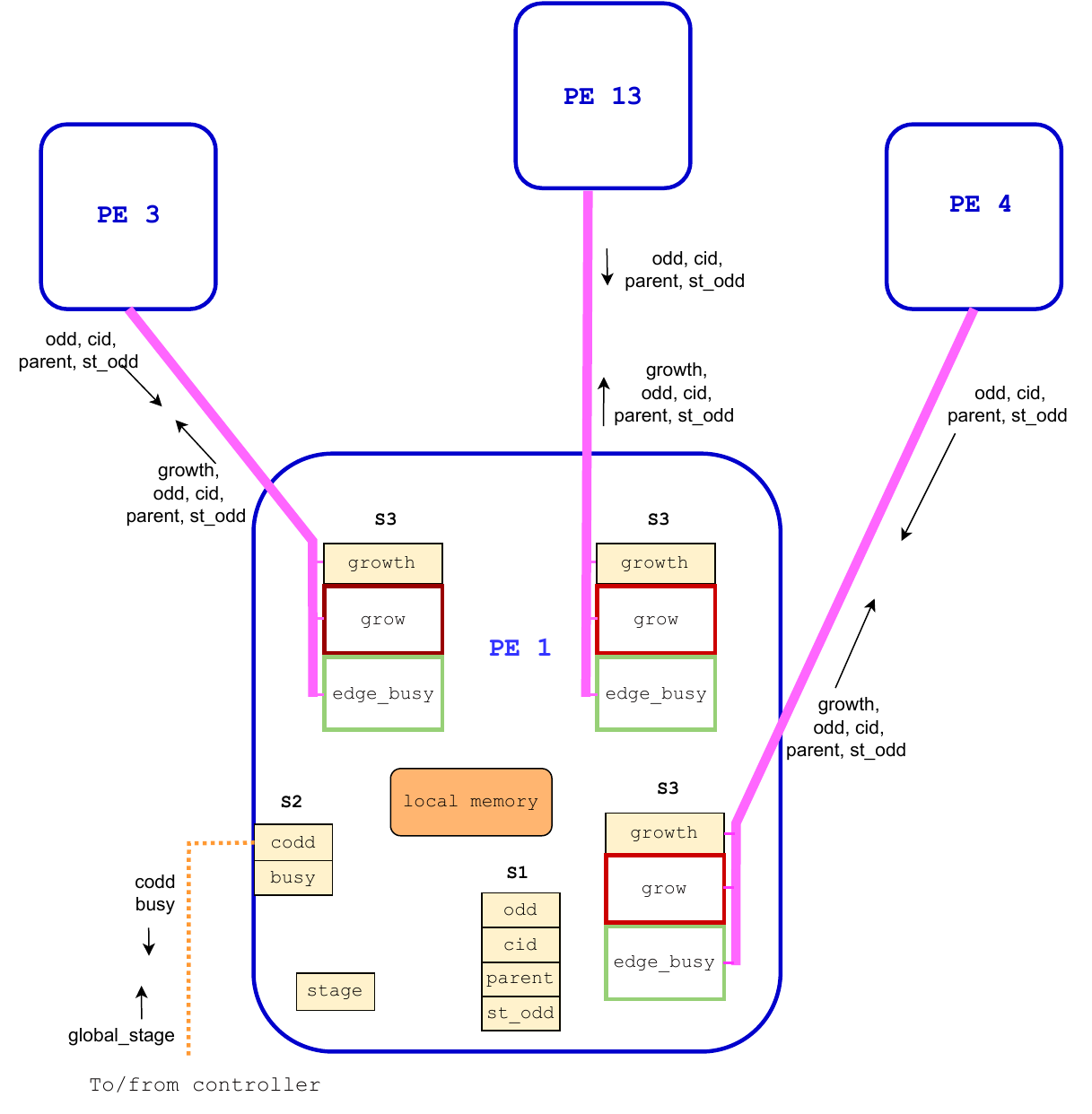}

\caption{The bottom left corner of the PE array shown in \autoref{fig:pe_array}. Only part of the logic and memory inside PE 1 is shown: \code{growth} (S3) is per edge and is stored in the PE with lower $id$. \code{grow} logic (in brown) calculates the updated growth value. \code{edge\_busy} (in green) is per adjacent PE and is used to calculate $v.$\code{busy}.
} 
  \label{fig:incidentPEs}
\end{minipage}
\end{figure*}

\subsection{Worst-Case Time Complexity Analysis}

The worst-case time complexity of our distributed UF decoder is no worse than $O(d^{4} \log(d))$, which is the product of the worst-case number of stages, $O(d^{4})$, and the worst-case time complexity of the controller to change stages, $O(\log(d))$.

We show the worst-case number of stages is no-worse than $O(d^{4})$ as follows.
The number of stages is bounded by the maximum number of \emph{Merging} and \emph{Checking} stages (\lineref{line:inner_do_start} to \lineref{line:inner_do_end}) per iteration of the \code{while} loop in \lineref{line:outer_while_loop}, times the number of iterations.
These stages in each iteration implement a shared memory based flooding and convergecast algorithm for all existing clusters in parallel~\cite{attiya2004distributed}. 
This algorithm has a worst-case time complexity of $O(d^3)$, where $d^3$ bounds the cluster size~\cite{attiya2004distributed}.
Because each stage implements a step in the flooding and convergecast algorithm, the maximum number of stages in each iteration is bounded by $O(d^3)$.

The number of iterations is bounded by $d$ as each iteration consists of a \emph{Growing} stage and the maximum number of iterations any cluster can grow is $d$.
Thus, the total number of stages is no-worse than $O(d^4)$.

The controller's time complexity is contingent upon the implementation of the shared memory for $v.\code{busy}$ and $v.\code{codd}$. 
Since both checks involve logical OR operations on individual PE information, the most efficient implementation consists of a logical tree of OR operations, yielding a time complexity of $O(\log(d))$.

Nevertheless, the worst-case scenario is extremely rare since larger clusters are exponentially less likely to occur. As shown in the empirical results reported in \secref{sec:results}, the average time grows sublinearly with $d$. 



%% file: dalgorithm.tex
\subsection{Overview}
Like the original UF decoder, our distributed UF decoder is also based on the decoding graph.
Logically, the distributed decoder associates a processing element (PE) with each vertex in the graph. Therefore, when describing the distributed decoder, we often use PE and vertex in an inter-exchangeable manner. 
All PEs run the same algorithm, specified by \autoref{alg:distributed_uf}.
Like the UF decoder, a PE iterates over the \emph{Growing} and \emph{Merging} stages with the \emph{Merging} split into two: \emph{Merging} and \emph{Checking}.
Within each stage, PEs operate independently. 
A central controller coordinates their transition from one stage to the next as specified by \autoref{alg:distributed_uf_gc}.

A key challenge to the PE algorithm is to (\textit{i}) merge clusters and (\textit{ii}) compute the cluster parity, \emph{without} central coordination.
To achieve (\textit{i}), each PE is assigned a unique identifier (a natural number) and maintains the identifier of the cluster it belongs to, $cid$. The $cid$ is the lowest identifier of all its PEs, and the PE of the lowest identifier is called the root of the cluster. When two PEs connected by a fully grown edge have different $cid$s, the PE with the higher $cid$ adopts the lower value, resulting in the merging of their clusters. 
To achieve (\textit{ii}), each PE maintains a parent.
When a PE adopts the $cid$ from an adjacent PE, it sets the latter as its parent.
The parenthood relation between PEs creates a spanning tree for each cluster that is maintained by PEs locally and in which every PE in the cluster has a directional path to the root of the cluster.
The cluster parity can be computed using a convergecast algorithm on the spanning tree. 
We describe the PE algorithm in detail in~\ref{sec:pe_algorithm}.


To implement our distributed UF algorithm, we require several PE states, some of which are located in shared memories. We limit all communication between PEs and between PEs and the controller to coherent shared memories to ensure fast communication and prevent stalling that could result from message-based communication. 

\subsection{PE States}
A PE has direct read access to its local states and some states of incident PEs.
A PE can only modify its local states.

Thanks to the decoding graph, a PE has immediate access to the following objects.

\begin{itemize}[topsep=0.3em,itemsep=0.02em, leftmargin=*]
    \item $v$, the vertex it is associated with.
    
    \item $v.E$, the set of edges incident to $v$.

    \item $v.U$, the set of vertices that are incident to any $e\in v.E$ other than $v$ itself. We say these vertices are adjacent to $v$.
\end{itemize}

The algorithm augments the data structures of each vertex and edge of the decoding graph, according to the UF decoder design~\cite{delfosse2017almost}. 
For each vertex $v\in V$, the following information is added

 \begin{itemize}[topsep=0.3em,itemsep=0.02em, leftmargin=*]

 \item  $id$ : a unique identity number which ranges from $1$ to $n$ where $n=|V|$. $id$ is statically assigned and never changes. 

 \item  $m$ is a binary state indicating whether the measurement outcome is a defect measurement (\code{true}) or not (\code{false}). $m$ is initialized according to the syndrome.

\item  $cid$: a unique integer identifier for the cluster to which $v$ belongs, and is equal to the lowest $id$ of all the vertices inside the cluster. The vertex with this lowest $id$ is called the cluster root. $cid$ is initialized to be $id$. That is, each vertex starts with its own single-vertex cluster. 
When $cid=id$, the vertex is a root of a cluster. 

\item $odd$ is a binary state indicating whether the cluster is odd. $odd$ is initialized to be $m$.

\item $codd$ is a copy of $odd$.

\item $\code{parent}$ is a reference to the parent. As noted before, this parenthood relationship creates a spanning tree that  connects all vertices (PEs) with directional edges.

\item $st\_odd$: a binary state representing the parity of $m$ of $v$ and all its descendants.

\item $\code{stage}$ indicates the stage the PE currently operates in 

\item $\code{busy}$ is a binary state indicating whether the PE has any pending operations.

\end{itemize}

\begin{algorithm}[!t]
\DontPrintSemicolon
\caption{Algorithm for vertex $v$ in the distributed UF decoder.}\label{alg:distributed_uf}
\setcounter{AlgoLine}{25}
\footnotesize

$v.cid \gets v.id$;
$v.odd  \gets v.m$;
$v.parent \gets v.id$;
$v.st\_odd \gets v.m$ \;
\While{true}{ \label{line:duf_outer_loop}
\If{$\code{global\_stage} = $\emph{\code{terminate}}}{
 \Return
}
    \textup{Wait until $\code{global\_stage} = ${\code{growing}}} \\
    growing($v$) \label{line:duf_grow_start} \\
    \textup{Wait until $\code{global\_stage} = ${\code{merging}}} \\
    \Do{ \label{line:duf_inner_loop}
    $\code{global\_stage} = $\emph{\code{merging}} \label{line:merge_again}}
    { \label{line:do_in_PE}
    merging($v$) \\
    \textup{Wait until $\code{global\_stage} = ${\code{checking}}} \\
    checking($v$) \\
    \textup{Wait until $\code{global\_stage} != ${\code{checking}}}
    }
}
\end{algorithm}

\begin{algorithm}[!t]
\DontPrintSemicolon
\caption{Vertex growing algorithm}\label{alg:growing_function}
\footnotesize
\setcounter{AlgoLine}{40}
\function{growing(vertex v)}{
    $v.\code{busy} \gets \code{true}$; 
    $v.\code{stage} \gets \code{growing}$ \label{line:grow_response}\\
    \If{$v.odd$}{ \label{line:duf_is_odd}
         \ForEachAtomic{\textup{$e=\langle u,v\rangle\in v.E$}}{
                \If{$e.$\emph{\code{growth}}$<e.w$ \textup{and} $u.cid\neq v.cid$}{ \label{line:comparenupdate} 
                    $e.$\code{growth}$\gets e.$\code{growth}$+1$ \label{line:grow}
                }
        }
    }
    $v.\code{busy} \gets \code{false}$; \\
}

\end{algorithm}

\begin{algorithm} [!t]
\DontPrintSemicolon
\caption{Vertex merging algorithm}\label{alg:merging_function}
\setcounter{AlgoLine}{51}
\footnotesize

\function{merging(vertex v)}{
$v.\code{busy} \gets \code{true}$; 
$v.\code{stage} \gets \code{merging}$ \label{line:merge_response}\\
~\\


        \ForEach{$u\in v.nb$}{
            \If{$u.cid<v.cid$}{ \label{line:check_cid}
                $v.cid \gets u.cid$  \label{line:set_cid}\\
                $v.parent \gets u.id$ \label{line:set_parent}
            }
        }
        ~\\
        $v.st\_odd \gets \text{XOR}(u.st\_odd| u \in v.\code{child}, m)$ \label{line:subtree_parity}\\
        ~\\
        
        \lIf{$v.parent = v.id$}{
            $v.odd \gets v.st\_odd$ \label{line:root_parity}
        }
        \lElse{
            \textup{$v.odd \gets u.odd$ where $v.parent = u.id$} \label{line:calc_odd}
        }
        ~\\
        $v.\code{busy} \gets \code{false}$
}
\end{algorithm}

\begin{algorithm} [!t]
\DontPrintSemicolon
\caption{Vertex checking algorithm}\label{alg:checking_function}
\setcounter{AlgoLine}{68}
\footnotesize

\function{checking(vertex v)}{
$v.\code{busy} \gets \code{true}$\\
~\\

\If{\textup{$ \forall u \in v.\code{nb}, (u.cid = v.cid$ \& $v.odd = u.odd)$ and $v.st\_odd = \text{XOR}(w.st\_odd| w \in v.\code{child}, m)$ and $(v.parent \neq  v.id$ or $v.odd = v.st\_odd )$ }} { \label{line:merge_busy_check}
            $v.\code{busy} \gets \code{false}$
        }

$v.\code{stage} \gets \code{checking}$ \label{line:check_response} \\
}

\end{algorithm}

\begin{algorithm} [!t]
\DontPrintSemicolon 
\setcounter{AlgoLine}{76}
\footnotesize
\caption{The controller coordinates all PEs along stages and detects the presence of odd clusters.}\label{alg:distributed_uf_gc}
\While{true}{ \label{line:outer_while_loop}
    $\code{global\_stage} \gets \code{growing}$ \label{line:grow_signal}\\
    Wait until $\forall v \in V, v.\code{stage} = \code{growing}$\\
    Wait until $\forall v \in V, v.\code{busy} = \code{false}$

    ~\\
    \Do{\textup{$\exists v \in V, v.busy = \code{true}$}}{ \label{line:inner_do_start}
    $\code{global\_stage} \gets \code{merging}$ \label{line:merge_signal}\\
    Wait until $\forall v \in V, v.\code{stage} = \code{merging}$\\
    Wait until $\forall v \in V, v.\code{busy} = \code{false}$ \\

    ~\\
    $\code{global\_stage} \gets \code{checking}$ \label{line:check_signal}\\
    Wait until $\forall v \in V, v.\code{stage} = \code{checking}$\\
    }\label{line:inner_do_end}

    ~\\
    \If{ \textup{$\forall v \in V, v.codd = \code{false}$} \label{line:check_odd}}{
        $\code{global\_stage} \gets \code{terminate}$ \label{line:terminate_signal}\\
        \Return\\
    }
}
\end{algorithm}

\noindent For each edge $e\in E$, the decoder maintains $e.$\code{growth}, which indicates the growth of the edge, in addition to $e.w$, the weight. $e.$\code{growth} is initialized as $0$. The decoder grows $e.$\code{growth} until it reaches $e.w$ and $e$ becomes \emph{fully grown}.

For clarity of exposition, we introduce a mathematical shorthand 
$v.$\code{nb}, the set of vertices connected with $v$ by full-grown edges, i.e., $v.$\code{nb}=$\{u|e=\langle v,u\rangle\in v.E~\land~e.$\code{growth}$=e.w\}$. We call these vertices the \emph{neighbors} of $v$. Note neighbors are always adjacent but not all adjacent vertices are neighbors.
We also use $v.$\code{child}, to indicate all child vertices of a vertex in the tree representation, i.e., $v.$\code{child}=$\{u|u.\code{parent} = v.id\}$. Since trees are built within a cluster, all child vertices are neighbors but not all neighbors are child vertices.

\subsection{Shared memory based communication}

We use coherent shared memory for a shared state that has a single writer.  For all shared memories, given the coherence, a read always returns the most recently written value. 
Like ordinary memory, we also assume both read and write are atomic.
\autoref{fig:incidentPEs} illustrates these memory blocks.

\begin{itemize} [topsep=0.3em,itemsep=0.02em, leftmargin=*]
    \item memory read/write for PE ($v$) and read-only for adjacent PEs, i.e., $\forall u\in v.U$. $v.id$, $v.cid$, $v.odd$, $v.parent$ and $v.st\_odd$ reside in this memory (S1).
    
    \item  memory read/write for PE ($v$) and read-only for the controller. 
    The PE local states,  $v.codd$, $v.$\code{stage} and $v.$\code{busy} reside in this memory (S2). 
    
    \item memory for  $e.$\code{growth}, which can be written by its two incident PEs (S3). 
    
    \item memory read/write for the controller and read-only for all PEs. The controller state \code{global\_stage} is stored in this memory (S4).
    
\end{itemize}

\subsection{PE Algorithm}
\label{sec:pe_algorithm}
All PEs iterate over three stages of operation. Within each stage, they operate independently but transit from one stage to the next when the controller updates \code{global\_stage}.
When a PE enters a stage, it sets $v.stage$ accordingly and keeps $v.\code{busy}$ as \code{true} until it finishes all work in the stage.
The controller uses these two pieces of information from all PEs to determine if a stage has started and completed, respectively (See \S\ref{sec:controller}).

We next describe the three stages of the PE algorithm.
In the \textbf{Growing} stage, vertices at the boundary of an odd cluster increase $e.$\code{growth} for boundary edges (\lineref{line:grow}).
As PEs perform Growing simultaneously, two adjacent PEs may compare $e.w$ and $e.growth$ and update $e.growth$ for the same $e$. 
Such compare-and-update operations must be atomic to avoid data race. 

In the \textbf{Merging} stage, two clusters connected through a fully-grown edge merge by adopting the lower cluster id ($cid$) of theirs.
To achieve this, each PE compares its $cid$ with its neighbors (\lineref{line:check_cid}).
If the other incident vertex of a fully grown edge has a lower $cid$, the PE adopts the lower $cid$ as its own (\lineref{line:set_cid}). 
The merging process continues until every PE in the cluster has the same $cid$, which is the lowest vertex identifier of the cluster.

In order to compute the cluster parity, when a PE adopts the $cid$ of the adjacent PE, it sets the latter as its \code{parent} (\lineref{line:set_parent}). 
This parenthood relation creates a spanning tree for each cluster that includes all PEs (vertices) with directional edges. 
Each PE then calculates the parity of itself and all its children as $st\_odd$ (\lineref{line:calc_odd}). 
Note that $odd$ of the root PE is the same as its $st\_odd$ (\lineref{line:root_parity}). 
All other PEs copy the $odd$ of their respective parents (\lineref{line:calc_odd}).

Astute readers may point out that $v.st\_odd$ should be the parity of $v$ and all its descendants, not just children.
This is achieved by two modifications, compared to the UF decoder.
First, a new stage \textbf{Checking} is added after Merging to see if the PE (vertex) needs to go back to \emph{Merging} again (\lineref{line:merge_busy_check}).
Second, all PEs iterates through Merging and Checking until all PEs have nothing to do for Merging. (\lineref{line:do_in_PE}-\lineref{line:merge_again}).
These allow parity computation to propagate from leaves to the roots of the spanning trees while $cid$ and $odd$ to propagate from the roots to the leaves. 

\paragraph{Building corrections within clusters}
While the original UF decoder builds a spanning tree within each even cluster in the end to generate a correction (\lineref{line:peeling}), 
our distributed UF decoder already has a spanning tree based on the parenthood relation and therefore is more efficient in generating corrections. 


\subsection{Controller Algorithm}
\label{sec:controller}


The controller moves all PEs and itself along the three stages. 
In the Growing and Merging stages, it checks for $v.\code{busy}$ signals from each PE. 
The controller determines the completion of a stage when all PEs have $v.\code{busy}$ as \code{false}.
In the Checking stage controller determines the completion of the stage when all PEs have moved to the Checking stage.
Upon completion, the controller updates the \code{global\_stage} variable to move to the next stage and the PEs acknowledge this update by updating their own $v.\code{stage}$ variable.


The controller also calculates the presence of odd clusters. 
At the end of the Merging and Checking stages, it reads the $v.odd$ value of each vertex (\lineref{line:check_odd}). 
If any vertex has $v.odd = true$, the controller updates the global stage variable to Growing to continue the algorithm. 
Otherwise, it updates it to Terminate to end the algorithm.

%% file: system.tex
\section{\name Architecture}
\label{sec:System}

We next describe \name, the architecture for the distributed UF decoder.
\subsection{Overview}

\name organizes PEs and the controller in a custom topology that combines a 3-D grid and a tree as illustrated by \autoref{fig:pe_array} and explained below.
\begin{itemize}[topsep=0.3em,itemsep=0.02em, leftmargin=*]
    \item PEs are organized according to the position of vertices in the model graph they represent. We assign $v.id$ sequentially, starting with 1 from the bottom left corner and continuing in row-major order for each measurement round. 
    Shared memory S1 ($v.cid$, $v.odd$, $v.parent$ and $v.st\_odd$) and S2 ($v.codd$, $v.$\code{stage}, and $v.$\code{busy}) are per PE. 

    \item Shared memory S3 ($e.\code{growth}$) is added to the incident PE with the lower $id$. 
    
    \item A link between every two adjacent PEs to read from each other's S1 and for the one with the higher $id$ to read the other's S4. This results in a network of links in a 3-D grid topology. As a PE represents a vertex in the model graph, a link represents an edge. Broad pink lines in \autoref{fig:pe_array} represent these links.

    \item The controller is realized as a tree of control nodes (\secref{ssec:coordination_tree}). 
    The leaf nodes of the tree contain shared memory S4.

    \item A link between each PE and the controller for the controller to read from S2 and for the PEs to read from S4. Dashed orange lines in \autoref{fig:pe_array} represent these links.

\end{itemize}

\subsection{Controller}
\label{ssec:coordination_tree}
\name implements the controller as a tree of control nodes to avoid the scalability bottleneck. 
The controller requires three pieces of information from each PE: $v.codd$, $v.\code{stage}$ and $v.\code{busy}$.
Each leaf control node of the tree is directly connected with a subset of PEs. 
We can consider these PEs as the children of the leaf node.
Each node in the tree gathers vertex information from its children and reports it to the parent.
With information from all vertices, the root control node runs ~\autoref{alg:distributed_uf_gc} and decides whether to advance the stage. 

We leave height, branching factor, and the subset of PEs connected to each leaf node as implementation choices. 
The necessary requirement is that the controller should not slow down the overall design.






%% file: implementation.tex
\section{FPGA Implementation}
\label{sec:implementation}

We next describe an implementation of \name targeting a single FPGA.
We choose a FPGA for two reasons. It supports massively parallel logic, which is essential as the number of PEs grows by $d^3$ in our distributed UF design.
Moreover, it allows deterministic latency for each operation, which facilitates synchronizing all the PEs.
Our implementation contains approximately 3000 lines of Verilog code, which is publicly available at~\cite{qecGithub}.

\subsection{Leveraging global synchronization in FPGA}
\label{ssec:time_complexity}

\begin{figure} [!t]
	\centering
\includegraphics[width=0.80\linewidth]{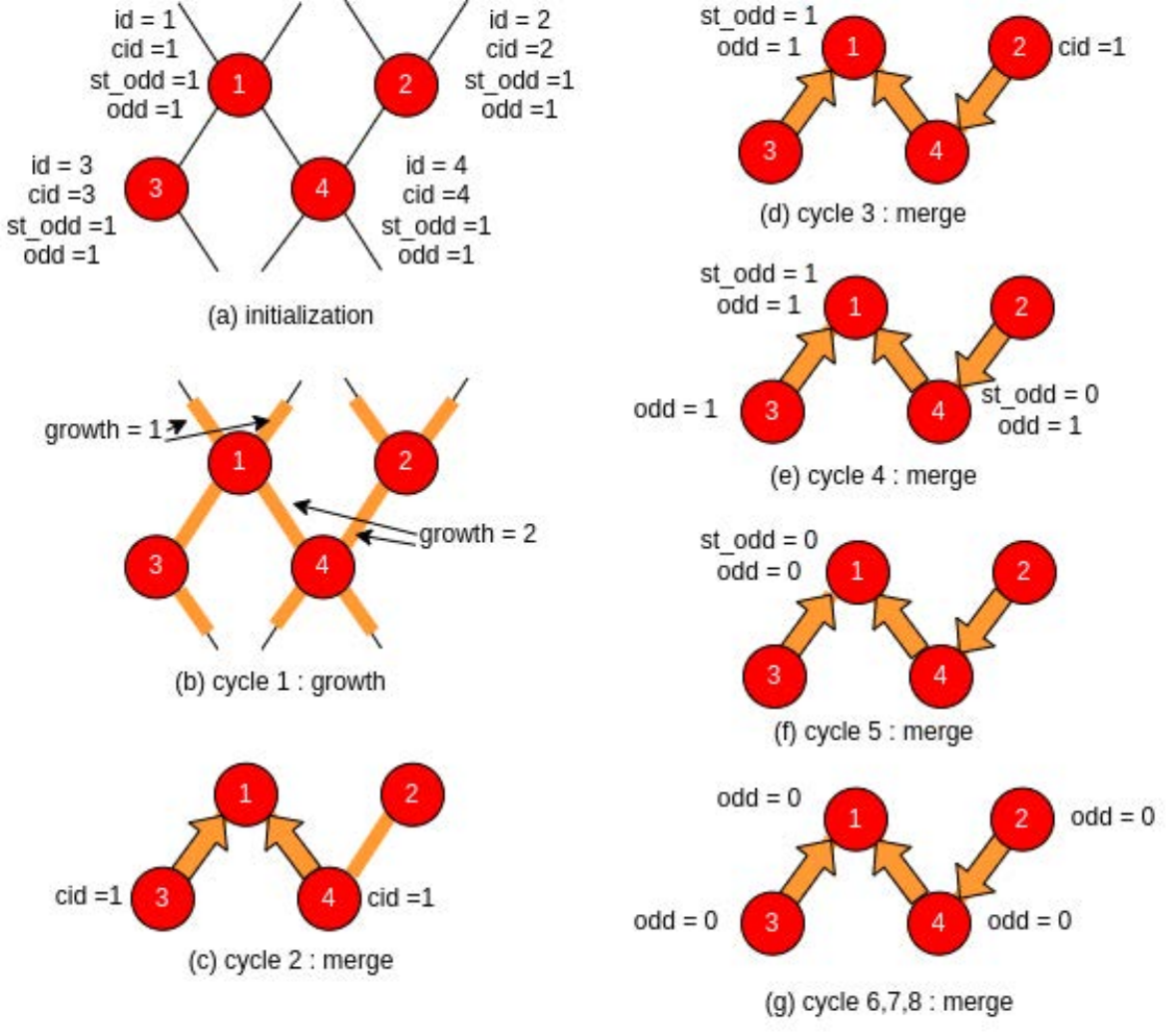}
\caption{An example figure showing how the FPGA implementation groups four nearby defect measurements into a single cluster in eight cycles. (a) Each defect measurement is mapped to a PE and initially, the four defect measurements have $v.id=1,2,3,4$, $v.cid = v.id$, $v.st\_odd=v.odd=1$. (b) The first growth cycle results in fully grown edges between \{1,3\}, \{1,4\} and \{2,4\}. (c) During merging, PEs 3 and 4 set their $v.cid$ as 1 and set their parents to 1 (shown with orange arrows). (d) In the next cycle, PE 1 calculates the parity of the subtree rooted at 1 (PEs 1, 3, 4) while PE 2 updates its $v.cid$ and parent. (e,f) This results in an update of $v.st\_odd$ of subtrees rooted at 4 and 1 in the next two cycles. Simultaneously, the root node (PE 1) updates the parity of the cluster ($v.odd = 0$). (g) $v.odd$ is propagated to all PEs in the cluster in two cycles, and no change occurring in the 8th cycle tells the controller to advance the stage.}
 
	\label{fig:cycle_diagram}
\end{figure}

We leverage global synchronization inside the FPGA to speed up our distributed UF algorithm. Running the FPGA design in a single-clock domain allows us to have all the PEs and the control nodes tightly synchronized. 
Notably, we simplify our algorithm as follows. 
Firstly, we run the Merging (\lineref{line:FPGA_merging}) and Checking stages (\lineref{line:FPGA_checking}) in parallel within each PE.
The tight synchronization of all PEs guarantees that false negative \code{busy} signals do not occur.

Secondly, we reduce the overhead of synchronization by having the controller only coordinate moving to the Growing stage at the beginning of each iteration (\lineref{line:PE_move_growing}). As each PE can perform the Growing stage deterministically in a single cycle, PEs can move to the Merging stage without central coordination (\lineref{line:PE_move_merging}). 

Additionally, as the controller deterministically knows the exact stage each PE is in, \code{stage} is stored locally and not shared with the controller. Thus the information from the PEs to the controller is limited to two bits, $v.\code{busy}$ and $v.\code{odd}$. 

\autoref{alg:FPGA_distributed_uf} and \autoref{alg:FPGA_gc} lists the FPGA-oriented algorithm of PE and the controller. The logic at every positive edge is executed in parallel. 
\autoref{fig:cycle_diagram} provides a simple example of how the FPGA implementation merges a cluster of four defect measurements in eight cycles.

\paragraph{Time Complexity}

The worst-case cycle count of the FPGA design is bounded by $3d^4 + 2d$. 
The merging stage consists of three primary operations: a broadcast ($cid$), a convergecast ($st\_odd$), and another broadcast ($odd$) with each operation requiring at most $d^3$ cycles. 
In addition, the merging stage needs an extra cycle to verify completion. 
Conversely, the growth stage requires a single cycle. 
As a result, each iteration requires at most $3d^3+2$ cycles. 
Since number of iterations is at most $d$, the worst-case cycle count is bounded by $3d^4 + 2d$.
The worst-case time complexity of the FPGA design is $O(d^{4})$ in contrast to $O(d^{4}\log(d))$ of the distributed UF algorithm.
The $\log(d)$ factor in the latter originates from the coordination overhead associated with transitioning between the stages. 
The FPGA design performs stage transition in a single cycle, effectively eliminating the $\log(d)$ factor.

 \subsection{Optimizing resource efficiency}

As the resource usage grows $O(d^3log(d))$, the number of LUTs limits the largest $d$ that can be implemented on a given FPGA. 
To address this constraint, we adopt a method first proposed by Heer et al.~\cite{Heer2023AchievingSQ}.
This method first partitions the decoding graph into multiple sub-graphs and then time-multiplexes them in the FPGA.

We first partition the decoding graph into multiple sub-graphs by splitting it evenly along one or more axes.
This even partitioning ensures that each sub-graph is roughly the same size, thereby increasing resource utilization.
The necessary condition for partitioning is that each sub-graph must be sized to fit within a single FPGA. 

Time multiplexing of multiple sub-graphs is as follows.
We first implement a graph in the FPGA with the same topology as a sub-graph and at least as large as the largest sub-graph, which we will call a \textit{lattice}. 
We then iteratively map each sub-graph to the lattice during each decoding stage.
All sub-graphs of the decoding graph can be mapped to the same lattice due to the homogeneous topology of the decoding graph where each PE has a fixed number of edges that connect to adjacent PEs.
If $n$ subgraphs timeshares a lattice, we denote the implementation as \name-$n$. By default, \name denotes \name-$1$ when there is no multiplexing.

We implement context-switching between sub-graphs at the PE level. 
We augment each PE in the lattice, which we label as a \emph{physical PE}, with a local memory.
During context-switching each PE stores its PE states in the local memory and loads the PE states of the corresponding PE of the next subgraph from the local memory.
This is akin to context switching of threads in an operating system.
\autoref{fig:incidentPEs} shows a minimal diagram of a physical PE.
In the FPGA, this local memory is mapped to LUTRAM rather than BRAM, due to its shallower depth.
We also note that context-switching in \name consumes a single cycle as it is essentially reading and writing from local memory.

In the example in \autoref{fig:cycle_diagram}, adding context-switching requires two additional cycles for the cluster, occurring after the growing stage (cycle 1) and the merging stage (cycle 8).


A careful reader may point out that time-multiplexing through multiple sub-graphs would require extra connections between physical PEs to provide adjacent PE information to virtual PEs mapped to the boundary of the lattice.
Indeed, this is the case for the original design proposed by Heer et al.~\cite{Heer2023AchievingSQ}.
We avoid these extra connections by carefully mapping virtual PEs to physical PEs.
We always map any pair of adjacent virtual PEs in the decoding graph belonging to different sub-graphs, to the same physical PE.
Thus, the missing information of adjacent PEs at the lattice boundary can be loaded from the local memory of the physical PE.




\subsection{Implementation details}

We next list the other implementation choices of our design.

\textit{Contetxt switching}:~~
On the VCU129 FPGA development board~\cite{vcu129}, without context switching we can support the distributed UF decoder with $d$ up to 21 for phenomenological noise model and up to 17 for circuit level noise, due to resource limits.
We use context switching only for $d$ exceeding those limits. 
Furthermore, we restrict the partitioning of the decoding graph along a single axis to avoid excessive sequential reads from local memory by physical PEs at the boundary of the lattice.

\textit{Controller}:~~
Since the largest number of PEs we can implement a single VCU129 FPGA is 4620 ($d=21$), a single node controller suffices.
The node controller reads \code{busy} of each PE, every clock cycle to identify the completion of a stage.

\begin{figure}[t]
    \centering
        \begin{subfigure}{0.49\linewidth}
	    \centering
        \includegraphics[width=1\textwidth]{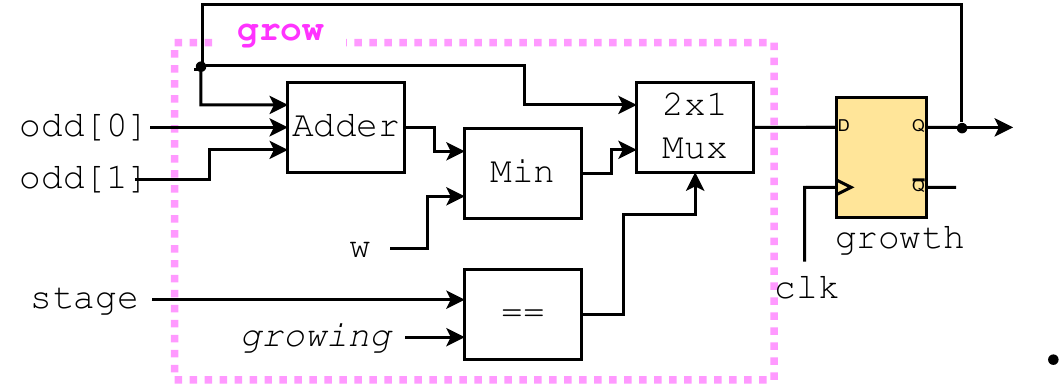}
        \label{}
        \end{subfigure}
    \begin{subfigure}{0.49\linewidth}
	    \centering
\begin{lstlisting}[language=Verilog, frame=single, basicstyle=\scriptsize\ttfamily]
reg growth;
always@(posedge clk)
  if(stage == growing)
    growth <= `MIN(growth 
      + odd[0] + odd[1], w);

\end{lstlisting}
\label{}
        \end{subfigure}
\caption{Circuit diagram of \code{grow} sub-module and Verilog implementation. This implements the  atomic compare and update operation in \lineref{line:comparenupdate} as part of the PE module. $odd[0]$ and $odd[1]$ represents the $odd$ state of the two incident PEs of the edge.}

\label{list:egrowth}
\end{figure}

\textit{Shared memory}:~~
We implement all shared memories as FPGA registers, i.e., \textbf{\code{reg}} in Verilog. 
FPGA registers by design guarantee that a read returns the last written value. 
In order to ensure that the S4 memory has a single writer, we adjust the PE logic to update growth by implementing a modified compare-and-update operation (\lineref{line:FPGA_comparenupdate}) as shown in~\autoref{list:egrowth}. The PE that houses the S3 memory performs this operation, increasing $e.\code{growth}$ by two when both endpoints of the edge have $v.odd$ set to true.

\subsection{Resource Usage}

\autoref{tab:FPGA_usage} shows the resource usage for various $d$ for phenomenological noise model and circuit-level noise.

While the numbers of vertices and edges grow by $O(d^3)$, resource usage grows faster for the following reasons. First, resource usage by a PE grows due to the increase of bit-width required for $v.id$, and $v.cid$. A PE for $d=21$ with six adjacent PEs requires 200 LUTs and a similar PE for $d=5$ requires only 155 LUTs.
Second, PEs on the surface of the three-dimensional array as shown in \autoref{fig:pe_array} use fewer resources than those inside because the latter have more incident edges. When $d$ increases a higher portion of PEs are inside the array.
The increased number of incident edges also causes the Helios for circuit-level noise to use twice the resources as in circuit-level noise each PE inside the graph can have up to 12 incident edges. 


Existing commercial FPGAs like VCU129 often dedicate a lot of silicon to digital signal processing (DSP) units and block RAMs (BRAMs). 
However, our design does not use any DSPs because it only requires comparison operators and fixed point additions.
We only use BRAMs to support interfacing with the MicroBlaze core and implement context switching using LUTRAMs instead of BRAMs
Therefore, an ideal FPGA designed to run our distributed UF decoder would be simpler than current large FPGAs, as it would only need a large number of LUTs, no DSP units, and a limited amount of BRAM.

\input{TABLE1}

\subsection{Clock Frequency}
\label{sec:frequency}


The architectural mismatch between the 3-D design of \name and the 2-D structure of an FPGA creates a fundamental limitation in the maximum clock frequency \name can run when implemented on an FPGA. 
Despite \name's capability to scale to arbitrarily large $d$, the maximum clock frequency of the FPGA implementation must decrease as $d$ increases and eventually \name will not be able to decode at the rate of measurement.
In our implementation, we reach the limit of FPGA resources before the clock frequency becomes the bottleneck, at $d \leq 51$.
We estimate that with an arbitrarily large FPGA with the same routing technology as Virtex UltraScale+ device, \name's clock frequency will become the bottleneck and will fail to decode at the rate of measurement at around $d\approx 1800$.

The signal propagation latency increases by $O(d)$ in the FPGA implementation, causing a decrease in maximum clock frequency as observed in ~\autoref{tab:FPGA_usage}.
When $d$ increases, PEs adjacent in \name must be placed farther apart within the FPGA, causing this increase.
Specifically, the critical path's routing latency increases from 3.77 ns for a $ d=3$ circuit-level noise model design to 11.56 ns for a $d=17$ design.
Additionally, the increase in bit-width of $d$ increases the logical processing latency by $O(log(log(d)))$, which is significantly less compared to the delay due to propagation.
When we synthesize PEs in isolation, the logic delay increases slightly from 1.475 ns to 1.552 ns when increasing $d$ from 5 to 21. 
We should also note that the significantly high maximum operating frequency for $d=3$ under phenomenological noise is due to the unique situation of the decoding graph for $d=3$, where no PE has more than three incident edges.


A potential approach to circumvent this architectural mismatch is to preserve \name's 3-D structure by mapping the decoder across multiple FPGAs.
However, the limitation of I/O pins in existing FPGAs and significant inter-FPGA latency of a few tens of nanoseconds prohibit practical implementation of \name across multiple FPGAs efficiently compared to a single FPGA implementation.


\textit{Implementation choice}:~~
For most experiments, we synthesize the design targeting a clock frequency of 100 MHz. 
This choice ensures sufficient latency for completing the critical path within a single clock cycle, allowing for a uniform comparison of the effects of our distributed union-find decoder.
We used slower clock frequencies, which were necessary due to resource congestion, only for the implementations of circuit-level noise at $d=17$ and the resource-efficient implementation at $d=51$.

\subsection{Power Consumption}

The power consumption of the implementation depends upon the number of PEs actively participating in the clustering process. 
For $d=13$, the Vivado synthesizer estimates power consumption at 4.639 W for the FPGA implementation.
This estimation is based on assuming random input values toggled continously, which results in all PEs being active at the same time~\cite{VivadoPower}. 
The power consumption during decoding is likely to be much lower
because most syndromes contain only a small number of defect measurements and as a result, only a small number of PEs are typically active at a time during the decoding process.

%% file: TABLE1.tex
\begin{table}[t]
\caption{Resource usage of \name on VCU129 FPGA board for selected $d$. \# LUTs and \# regs show the number of LUTs and registers required for each configuration, and $f_{max}$ shows, in MHz, the maximum clock frequency each configuration can run. }
\label{tab:FPGA_usage}
\small
\begin{center}
\begin{tabular}{|c|r|r|r|r|r|r|} \hline

\multirow{2}{*}{$d$} & \multicolumn{3}{c|}{phenomenological} & \multicolumn{3}{c|}{circuit-level} \\ \cline{2-7}
                                &  \# LUTs & \# regs & $f_{max}$ & \# LUTs & \# regs  & $f_{max}$ \\
 \hline\hline
3        &  970    &  528   & 260 & 1557 & 699  & 195     \\ \hline
5        &  6425    &   2425 & 220 &11128 & 3469 & 130   \\ \hline
9        &  52111    & 13754 & 175  & 93797 & 22515 & 110    \\ \hline
13        &  165718    &   47211 & 140  & 340084 & 74927 & 100    \\ \hline
17        &  448314 & 122028 & 125 & 888854 & 177284 & 75\\ \hline
21        &  898715 & 238939 & 100 & n/a & n/a & n/a \\ \hline
\end{tabular}

\end{center}
\end{table}

%% file: results.tex
\section{Evaluation}
\label{sec:results}

The main objective of our evaluation is to assess the scalability of our distributed UF implementation. To that end, we answer the following questions in our evaluation
\begin{itemize}
    \item \textbf{Latency growth}: Does the latency of distributed-UF decoder grow sub-linearly for both phenomenological noise and circuit level noise?
    \item \textbf{Context Switching overhead}: Can we use context-switching to decode large surface codes without excessive latency growth?
    \item \textbf{Extensibility}: Can Helios architecture be extended to support erasure errors, weighted edges, and sliding-window decoding?
\end{itemize}

We first describe our methodology and follow that with the evaluation results to answer the above questions.



\subsection{Methodology}
\begin{figure} [!t]
	\centering
\includegraphics[width=0.80\linewidth]{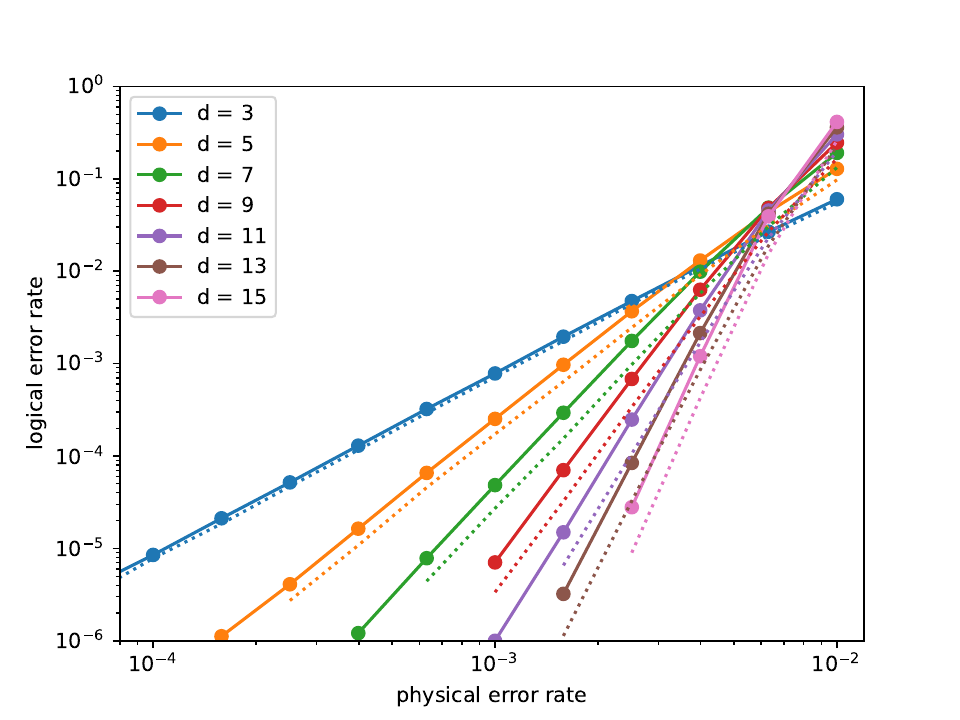}
\caption{Logical error rate of distributed UF decoder (dark lines) in comparison with MWPM decoder (dashed lines).}
 
	\label{fig:accuracy}
\end{figure}

For speed, we measure the number of cycles required to decode a syndrome. 
To evaluate correctness, we compare the results of our distributed UF decoder with those of the original UF decoder.
We compare clusters because the original UF decoder and ours only differ in implementing clustering.
In the rest of our evaluation, we will focus only on the speed of the distributed UF decoder and not on the accuracy of its results as our decoder and the original UF decoder by Delfosse et.al~\cite{delfosse2017almost} produces the same decoding output for any given syndrome.
Nevertheless, for completeness, \autoref{fig:accuracy} compares the logical error rates between our distributed UF decoder and the Minimum Weight Perfect Matching (MWPM) decoder.
We obtain results in ~\autoref{fig:accuracy} using a software implementation of the UF decoder and the MWPM decoder under circuit-level noise, with each data point representing the average of $10^8$ trials~\cite{qecPlayground}.

\paragraph{Experimental Setup}
As our evaluation setup, we use Xilinx VCU129 FPGA development board~\cite{vcu129}, which contains one of the largest FPGAs available on a Xilinx development board.
We simulate a surface code on a PC under various noise models to generate syndromes, storing the output in a file. 
Subsequently, a MicroBlaze soft processor core~\cite{microblaze}, instantiated within the FPGA, reads this syndrome file. The core then transmits the syndromes to Helios, which operates within the same FPGA.
We ran $10^6$ trials for each error rate and distance.

\paragraph{Noise Model}
We use phenomenological noise model~\cite{dennis2002topological}, circuit-level noise model~\cite{landahl2011faulttolerant}, and phenomenological noise model with erasure errors~\cite{delfosse2017almost}. 
As decoding for X-errors and Z-errors are independent and identical, we only focus on decoding X-errors in the evaluation.

We use three noise models in our experiments: the phenomenological noise model~\cite{dennis2002topological}, the circuit-level noise model~\cite{landahl2011faulttolerant, qecPlayground}, and the phenomenological noise model with erasure errors~\cite{delfosse2017almost}. 
Each of these models additionally includes measurement errors. 
As the decoding for X-errors and Z-errors are independent and identical, we focus solely on decoding X-errors in our evaluation.

To simulate noise, we independently flip data qubits and ancilla qubits in our simulation model. 
In the phenomenological noise model, data qubits are independently flipped between each measurement round with a probability $p$.
For circuit-level noise, we flip both data and ancilla qubits between each pair of gates and between gates and measurements, also with a probability $p$. 
For erasure errors, we erase data qubits between measurement rounds with a probability $p_e$, and ancilla qubits adjacent to the erased qubit are flipped with a 50\% chance to emulate erasure effects. 
To emulate measurement errors we flip ancilla qubits with a probability of $p$.
This is a widely used approach by prior QEC decoders ~\cite{das2022afs, Holmes2020Nisq, Skoric2022Parallel, landahl2011faulttolerant, delfosse2017almost, wu2023Fusion}. 
We then generate the syndrome from the physical errors and provide it as input to our decoder.

For most of our experiments, we use as default $p=0.001$, like other works~\cite{das2022afs, wu2023Fusion, barber2023realtime}. 
This value is reasonable for surface codes, as $p$ should be sufficiently below the threshold (at least ten times lower) to exponentially reduce errors.
We note that the UF decoder has a threshold of $p=0.024$ for phenomenological noise calculated by Delfosse and Nickerson~\cite{delfosse2017almost}.
Similarly, for circuit-level noise the UF decoder has a threshold of $p=0.0078$ calculated by Barber et al.~\cite{barber2023realtime}.



\subsection{Decoding Time}
\label{sec:decodingtime}
We experimentally show how the average decoding time grows with the surface code size for different noise models.

\paragraph{Average time}

\begin{figure*}[!t]
        
	    \centering
        \begin{subfigure}{0.4\linewidth}
	    \centering
        \includegraphics[width=1.0\textwidth]{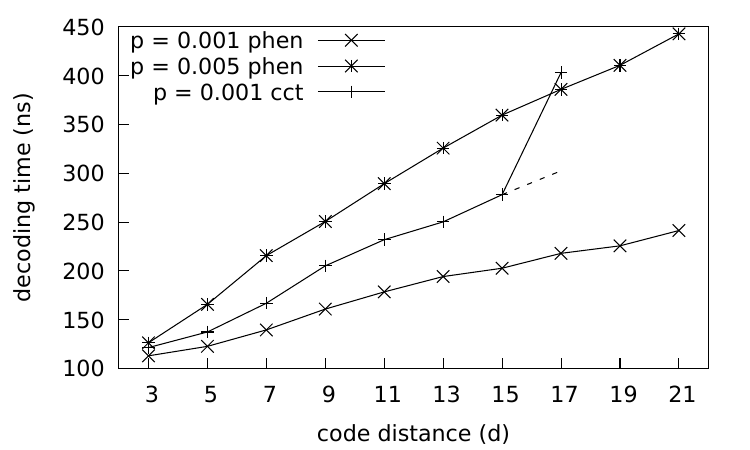}
        \caption{Average decoding time}
        \label{fig:d_growth}
        \end{subfigure}
        \hspace{5ex}
        \begin{subfigure}{0.4\linewidth}
	    \centering
        \includegraphics[width=1.0\textwidth]{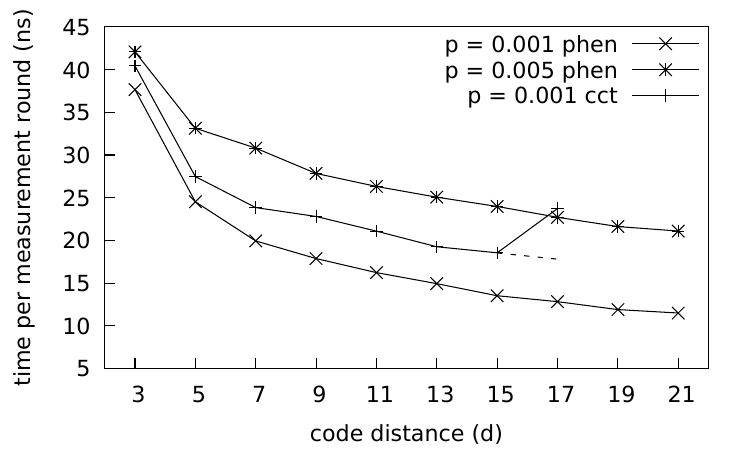}
        \caption{Average decoding time per measurement round.}
        \label{fig:per_d_rounds}
        \end{subfigure}

	\caption{Average decoding time scales sub-linearly with $d$. We measure the average decoding time for phenomenological noise ($phen$) of 0.005 and 0.001 and circuit level noise ($cct$) of 0.001. (Left) The average decoding time. The average time per measurement round reducing continuously justifies that our decoder is scalable for large surface codes under both phenomenological noise and circuit level noise. The unusual increase at $d=17$ for circuit level noise is caused by reducing the operating frequency to 75 MHz. The dashed line shows the calculated value at 100MHz. We show the distributions separately in \autoref{fig:d_variation}} \label{fig:DUF_decoding_time}
 \end{figure*}

To demonstrate the scalability of our algorithm with respect to the size of the surface code, we measure the average time for decoding for various sizes of the surface code. 
\autoref{fig:d_growth} shows the average decoding time in nanoseconds grows sublinearly with the distance ($d$) of the surface code (x-axis).
We see that for both phenomenological noise and circuit level noise we tested against, average decoding time grows sub-linearly with respect to the surface code size, which satisfies the scalability criteria to avoid an exponential backlog.
This implies that the average time to decode a measurement round reduces with increasing $d$ as shown in \autoref{fig:per_d_rounds}.



\paragraph{Distribution of decoding time}

\begin{figure}[!t]
	    \centering
     \begin{subfigure}{0.64\linewidth}
         \centering
        \includegraphics[width=1.0\textwidth]{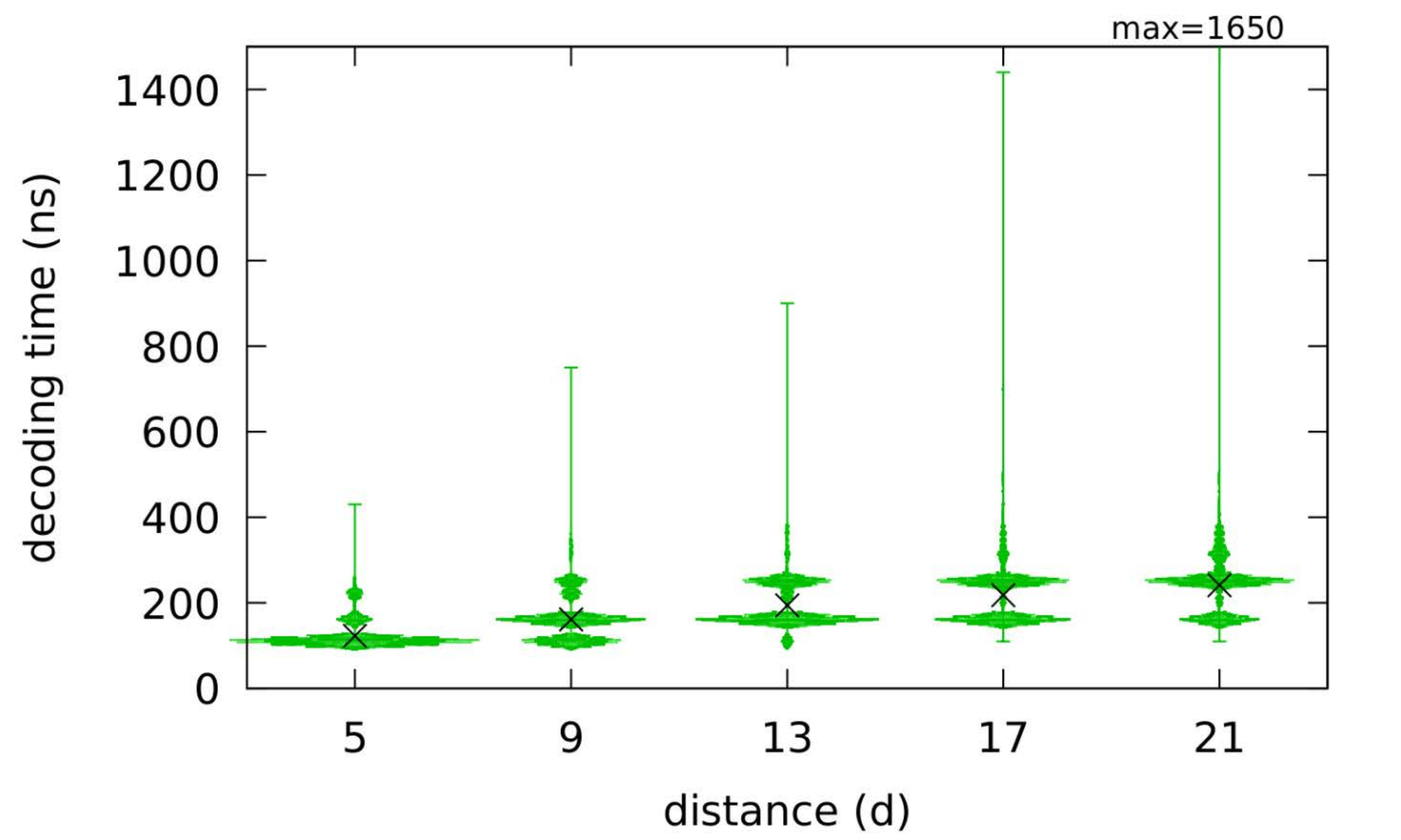}
	\caption{$T$ increases with $d$}
 	\label{fig:d_variation}
\end{subfigure}
\hfill
\begin{subfigure}{0.64\linewidth}	    
\centering
     \includegraphics[width=1.0\textwidth]{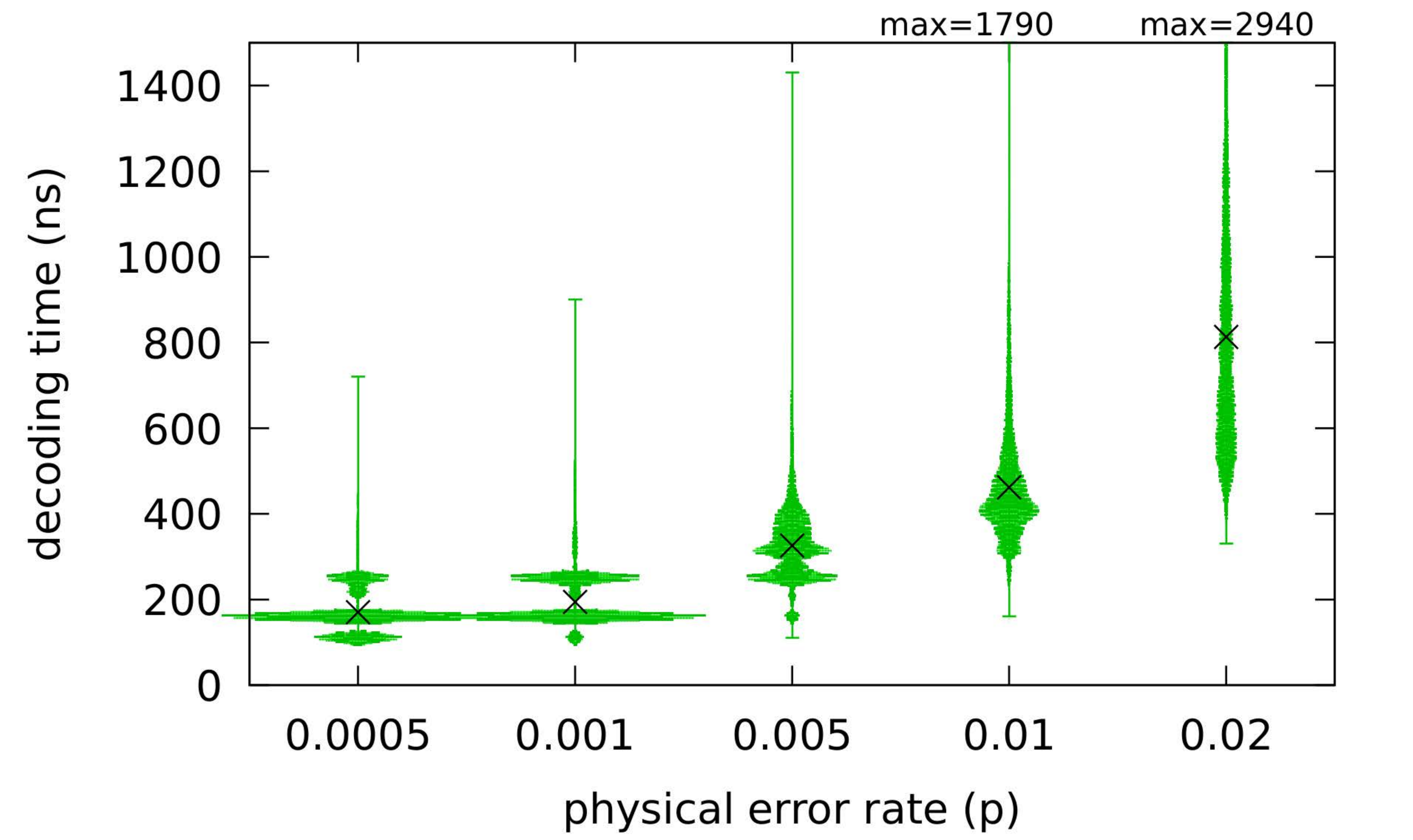}
	\caption{$T$ grows with the physical error rate.
 }
	\label{fig:noise_variation}
 \end{subfigure}
 \caption{Distribution of decoding time ($T$) with the mean marked with $\times$. Each distribution includes $10^6$ data points. By default $d=13$, phenomenological noise of $p=0.001$ and is unweighted }
  \end{figure}

To understand the growth of decoding time with respect to the code distance, in \autoref{fig:d_variation} we plot the distribution of decoding time for different code distances.
The y-axis shows the decoding time and the x-axis shows the distance ($d$) of the surface code.
We indicate the average cycle count with $\times$. 

  \begin{figure} [!t]
	\centering
\includegraphics[width=0.95\linewidth]{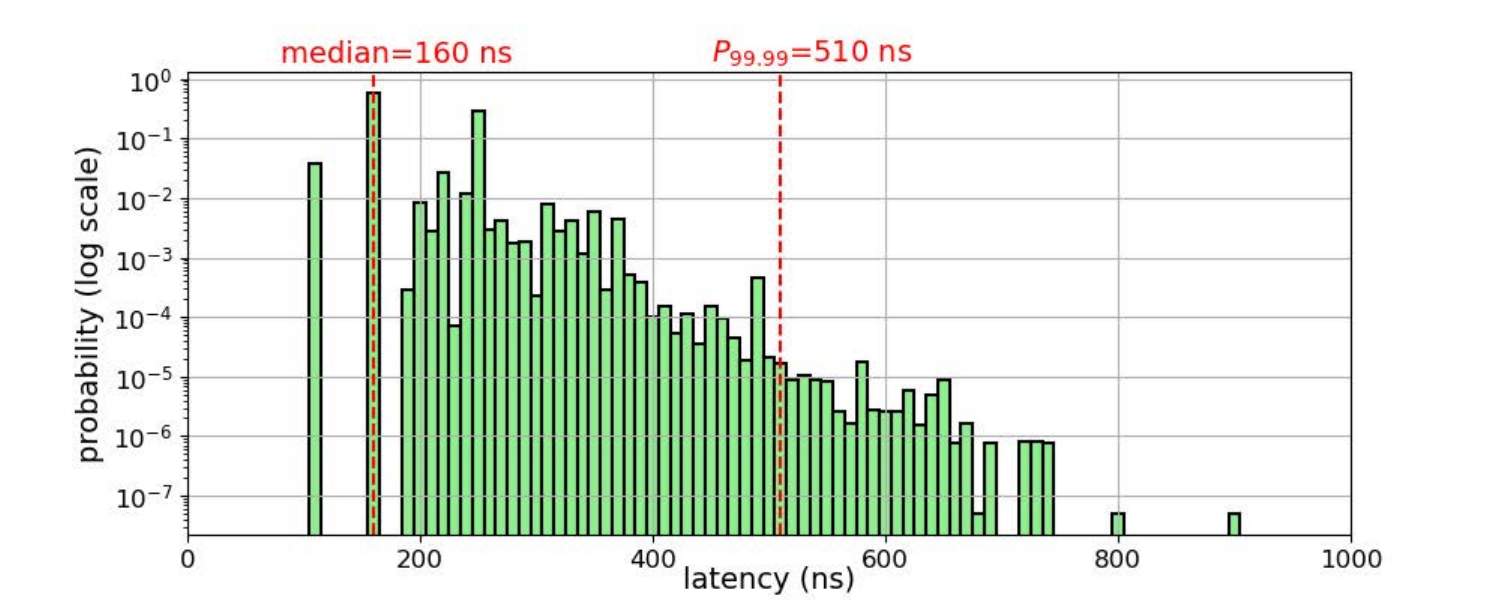}
\caption{Histogram of decoding time (latency) from $2\times10^7$ data points at $d=13$ and $p=0.001$ phenomenological noise.
This shows long decoding times are exponentially unlikely}
	\label{fig:log_distrib}
\end{figure}

The key factor determining the decoding time is the number of iterations of growing and merging the distributed UF decoder requires. The peaks in the probability distribution for each distance in \autoref{fig:d_variation} correspond to the number of iterations. The variation around each peak is caused by the time required to sync $c\_id$ and calculate $odd$.  The number of iterations is related to the size of the largest cluster, which in turn correlates with the size of the longest error chain in the syndrome. As the size of the surface code increases, the probability of a longer error chain also increases, resulting in the probability distribution shifting to the right. 

Furthermore, as seen in \autoref{fig:d_variation}, the distribution for each surface code size is right-skewed. 
For $d=13$, as seen in~\autoref{fig:log_distrib} $97\%$ of trials required two iterations or fewer, which were completed within 250 ns. In the same test, $99.99\%$ of trials were completed within 510 ns. 
Only exponentially fewer error patterns require long decoding times, corresponding to syndromes with longer error chains, which contributes less to the average decoding time. 
For example, excluding the $0.01\%$ samples in the tail yields an average decoding time of 194.16 ns, compared to 194.19 ns for all samples.

The longest decoding latency we observe in our trials, 920 ns, is significantly lower than the theoretical worse case decoding time of 857 $\mu$s, calculated by the equation in ~\autoref{ssec:time_complexity}. 
This discrepancy arises because the worst-case scenario requires a very specific pattern of syndromes, which is exceedingly rare and highly unlikely to occur in typical simulation settings.

\paragraph{Effect of physical error rate}

To understand the effect of the physical error rate on decoding time, in \autoref{fig:noise_variation} we plot the distribution of latency for five different noise levels for $d=13$. The y-axis shows the latency and the x-axis the physical error rate.

As the noise level increases, the probability distribution of latency shifts to the right. This is caused by the increased probability of a longer error chain when the physical error rate increases, which in turn requires more iterations to decode. As a result, the average decoding time increases with the physical error rate. 
For the highest tested physical error rate of 0.02, the average decoding time is 814.6 ns. 
Thus, even when the physical error rate is closer to the threshold, the decoder is an order of magnitude faster than the rate of measurement.

\subsection{Effect of optimizing for resource usage}
\label{sec:eval_resource}


\begin{figure*}[!t]
        
	    \centering
        \begin{subfigure}{0.32\linewidth}
	    \centering
        \includegraphics[width=1.0\linewidth]{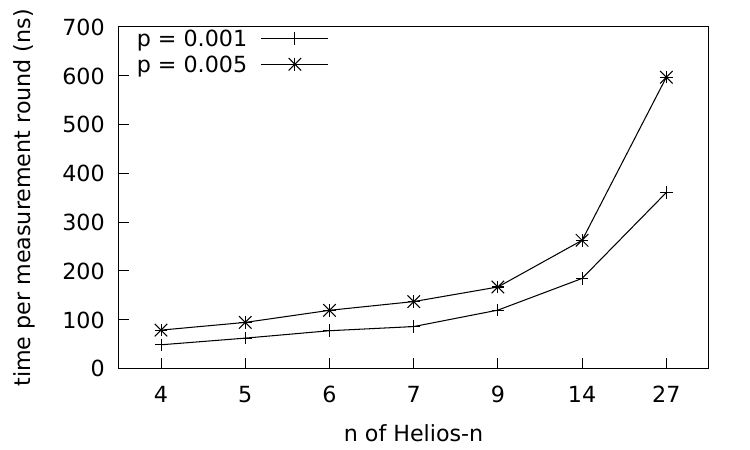}
        \caption{Latency increases with n}
        \label{fig:ctx_27_latency}
        \end{subfigure}
        \begin{subfigure}{0.32\linewidth}
	    \centering
        \includegraphics[width=1.0\linewidth]{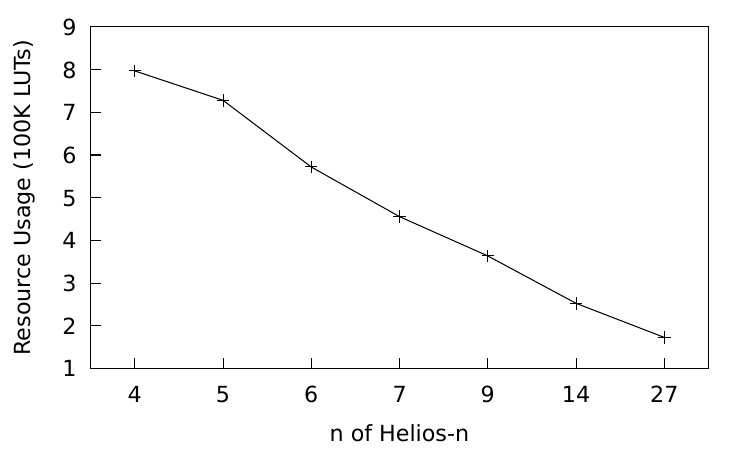}
        \caption{Resource use decreases with n}
        \label{fig:ctx_27_ru}
        \end{subfigure}
        \begin{subfigure}{0.32\linewidth}
	    \centering
        \includegraphics[width=1.0\textwidth]{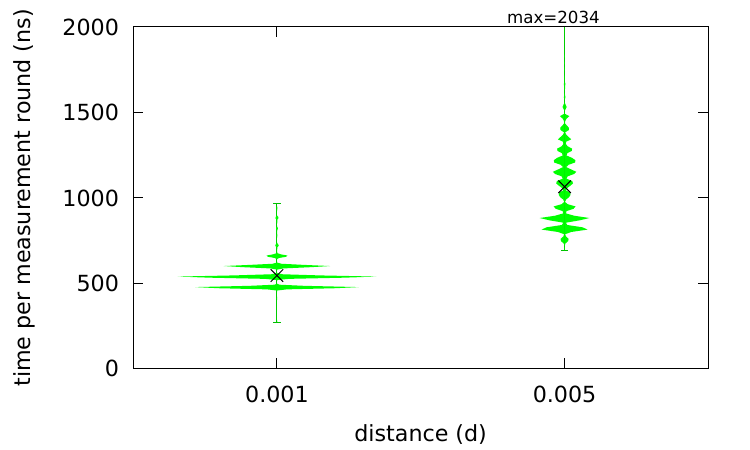}
        \caption{\name can decode $d=51$}
        \label{fig:ctx_51}
        \end{subfigure}

	\caption{\name can optimize for resource usage by mapping multiple virtual PEs to a single physical PE. (a) Average latency per measurement round for $d=27$ under two different phenomenological noise levels. (b) The corresponding resource use for the Helios-n configurations for $d=27$. (c) Distribution of decoding time for $d=51$ with \name-$51$. This configuration can decode faster than the rate of measurement for $p=0.001$, but is slightly slower than the rate of measurement for $p=0.005$}
 \end{figure*}

We next show that \name can decode surface codes larger than $d=21$ by dividing and time-multiplexing the decoding graph.
We first show that \name can decode $d=27$, a possible $d$ to run useful quantum algorithms~\cite{Gidney2021rsa} and then show \name can even decode significantly large $d$ such as 51.

In \autoref{fig:ctx_27_latency} we plot the average latency for decoding $d=27$ surface code under phenomenological noise of $0.001$ and $0.005$ for \name-$4$ to \name-$27$.
The Y-axis shows the average latency and the X-axis shows the number of LUTs required for implementation.
We only show the average as it is the critical factor enabling backlog-free decoding and the distribution observes a similar pattern as distributions shown in ~\autoref{fig:ctx_51}.
In \autoref{fig:ctx_27_ru} we plot the corresponding LUT count for Helios-n configurations shown in \autoref{fig:ctx_27_latency}.
We use the LUT count to indicate resource usage because it is the limiting factor when running a decoder on a given FPGA.
In \autoref{fig:ctx_27_latency} and \autoref{fig:ctx_27_ru}, we select Helios-n configurations resulting in maximum resource utilization.
Due to the restriction of partitioning solely across the measurement round axis, certain mappings, like \name-8, are inefficient. 
\name-8 will have a lattice with a height of four physical PEs, but the PEs in the topmost layer would only have three virtual PEs mapped to each physical PE. 
Conversely, \name-7 results in the same number of physical PEs but with a lesser number of context-switchings.

\autoref{fig:ctx_27_latency} shows that under phenomenological noise of $0.001$, \name can decode a $d=27$ surface code at an average latency of 48.5 ns per measurement round by mapping four virtual PEs to each physical PE. 
This rate is over 20 times faster than the measurement rate. 
By mapping 27 virtual PEs to each physical PE, resource usage can be further reduced to 173K LUTs, while maintaining the ability to decode at 360 ns per measurement round. 
In this configuration, each physical PE cycles through a single measurement result of the corresponding ancilla in each context.
The reduced LUT count is particularly significant as it enables the implementation to be mapped onto more cost-effective FPGA models.

\name can decode $d=51$ faster than the rate of measurement for phenomenological noise of $p=0.001$ but is slightly slower than the rate of measurement when $p=0.005$.
The average latencies for the noise level above are  543.9 ns and 1064.0 ns respectively.
This design targeting $d=51$ required around 796K LUTs and operates at 85 MHz.
Increased resource utilization at $d=51$, causes the reduction in operating frequency.
The distribution of latency is shown in \autoref{fig:ctx_51}.

\subsection{Decoder extensions}

\begin{figure*}[!t]
	    \centering
\begin{subfigure}{0.32\linewidth}	    
\centering
     \includegraphics[width=1.0\textwidth]{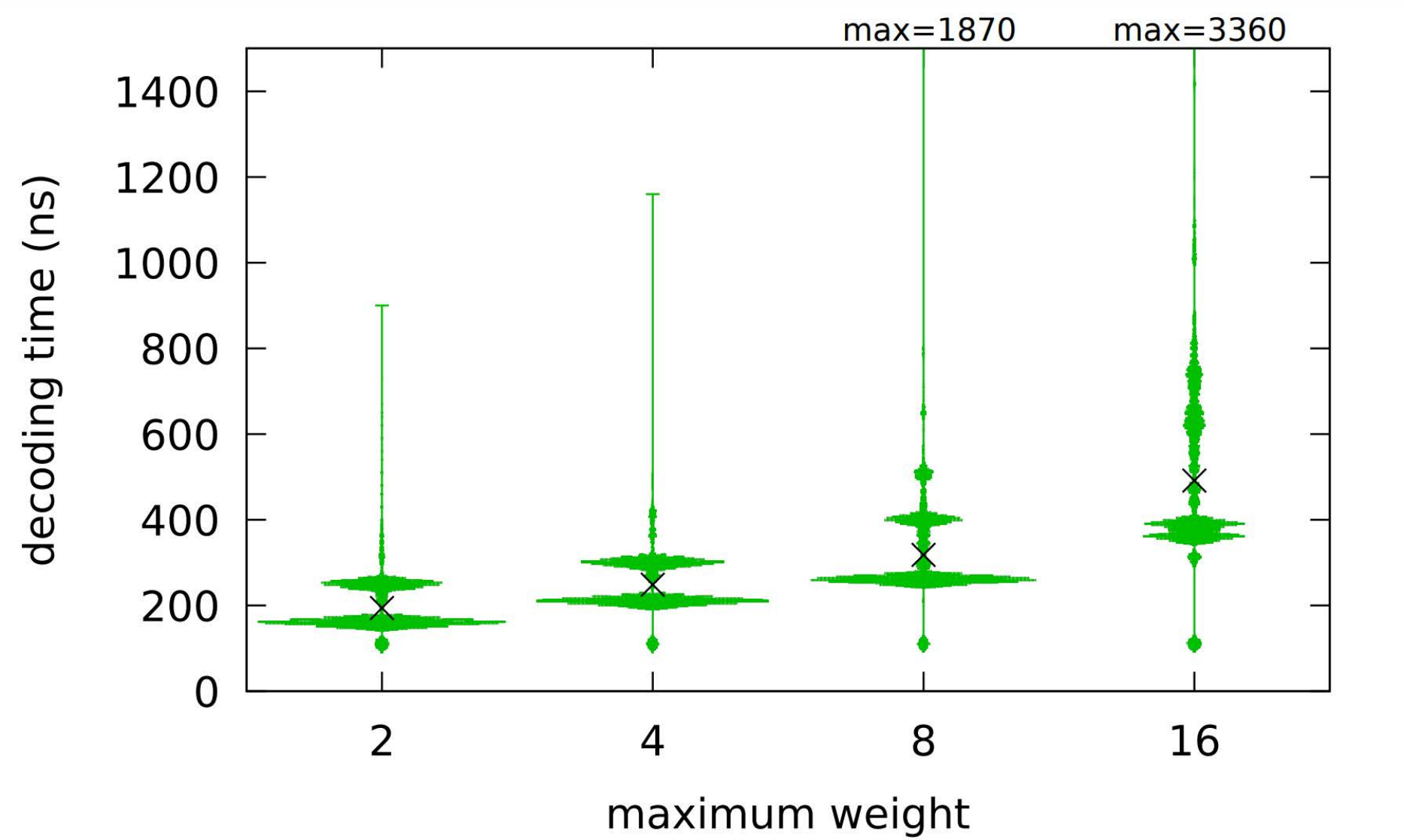}
	\caption{$T$ grows with the weight of the edges.
 }
	\label{fig:w_variation}
 \end{subfigure}
  \hfill
\begin{subfigure}{0.32\linewidth}	    
\centering
     \includegraphics[width=1.0\textwidth]{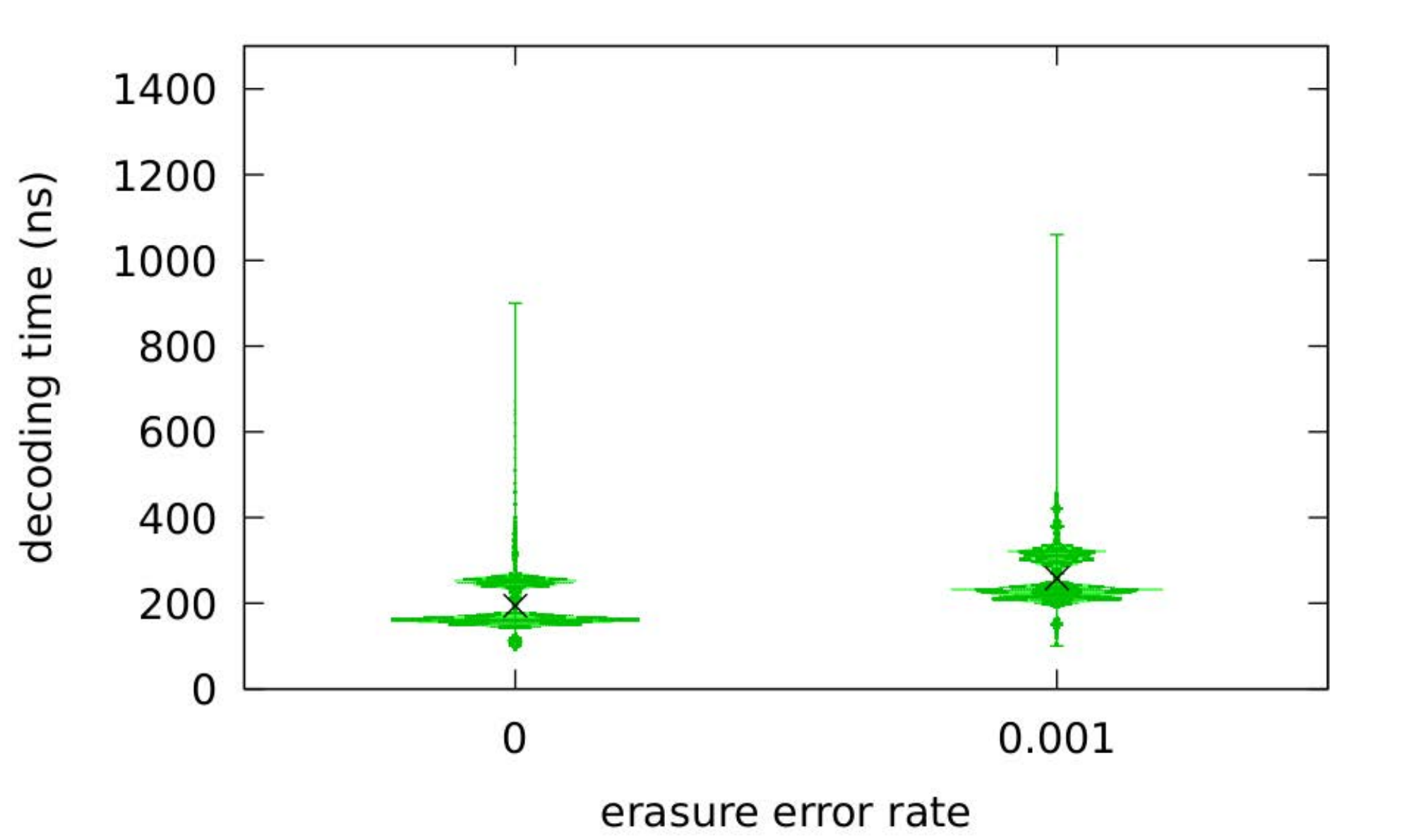}
	\caption{$T$ shifts with erasures.
 }
	\label{fig:e_variation}
 \end{subfigure}
\begin{subfigure}{0.32\linewidth}	    
\centering
     \includegraphics[width=1.0\textwidth]{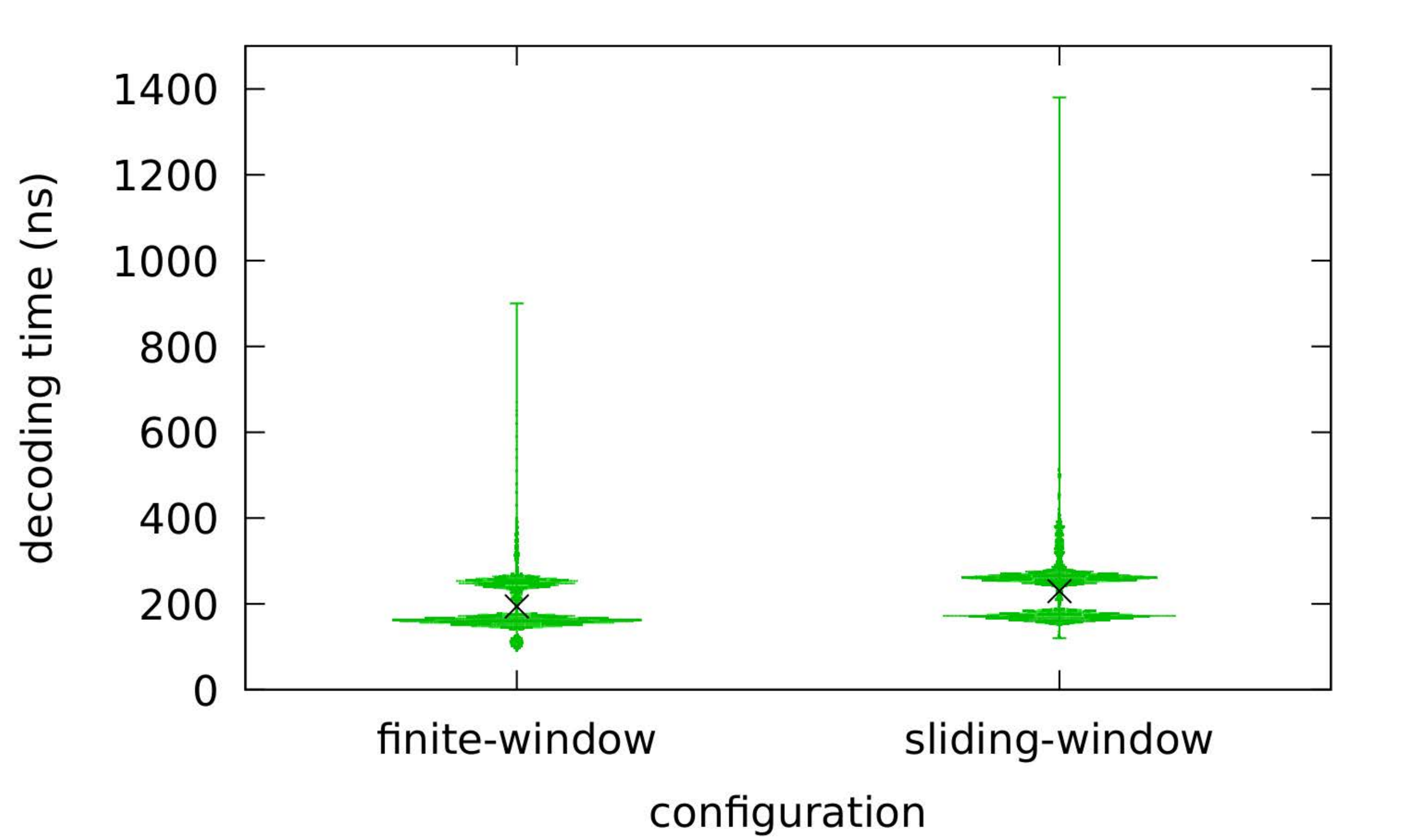}
	\caption{$T$ increases with sliding-window.
 }
	\label{fig:sliding_variation}
 \end{subfigure}
 \hfill
 \caption{Distribution of decoding time ($T$) for decoder extensions. The mean is marked with $\times$. Each distribution includes $10^6$ data points. By default $d=13$, phenomenological noise of $p=0.001$ and is unweighted }
  \end{figure*}

We next analyze the impact of extending our decoder for other requirements. We consider three situations: non-identically distributed errors, erasure errors and indefinite preserving of logical state.

\subsubsection{Non-identically distributed errors}

We next analyze the decoding process of a surface code with varying error probabilities for data and measurement qubits. While identically distributed errors are useful for evaluating the decoder's performance, practical implementation of surface codes may have different error probabilities for each qubit. To address this issue, each edge $i$ in the decoding graph is assigned a weight $w_i$ that ranges from 2 to $w_{max}$ and is proportional to $-\log(p_i)$, where $p_i$ is the error probability corresponding to edge $i$. $w_{max}$ is a user-specified parameter indicating the resolution of error probabilities.

\textbf{Noise model :} 
We assign random error probabilities from a standard normal distribution with a mean of $0.001$ and a standard deviation of $0.0005$.

 ~\autoref{fig:w_variation} shows that the average latency increases as $w_{max}$ increases. When the errors have a higher resolution, more iterations are required for each cluster, leading to an increase in latency. For the unweighted graph with $d=13$, the average decoding time per round of 15 ns  increases to 38 ns when $w_{max}$ increases to 16. Notably, all of these values are significantly faster than the rate of measurement. As a result, decoding non-identically distributed errors can be performed in real time using distributed UF on Helios.

 \subsubsection{Erasure errors}
The introduction of erasure errors slightly increases the decoding latency. 
 ~\autoref{fig:sliding_variation} shows the distribution of latency when erasure errors are added on top of $p$.
An erasure rate of 0.001 results in an increase in average decoding time by approximately 63 ns. 
Notably, 40 ns of this increase (4 FPGA clock cycles) comes from an extra \code{merging} stage prior to the initial \code{growing} stage.
This extra \code{merging} stage is necessary because erasure errors reduce an edge's weight to zero when an erasure occurs, which can lead to the merging of vertices before any growing of clusters.
Thus, the overall latency distribution is right-shifted by 40ns with a slight additional increase of latency due to X-errors caused by erasures. 

 
\subsubsection{Preserving logical state indefinitely}

We next show that \name can be extended to preserve logical state indefinitely using sliding window method~\cite{dennis2002topological}. 
While standard decoder evaluations focus on decoding $d$ rounds, practical applications require a decoder to continuously process incoming measurement rounds to maintain the logical state indefinitely. 
 The prevalent method for achieving this is the sliding window method. 
 In this method, $2d$ rounds are decoded simultaneously, but corrections are only committed for the oldest $d$ rounds. 
 Subsequently, the decoding window advances by $d$ rounds, resulting in continuous decoding for an indefinite period.

We implement sliding window decoding by extending the PE array for $2d$ measurement rounds.
This results in a slight increase in latency and more than a doubling of resource usage. 
For instance, a $d=13$ decoder supporting the sliding window method requires 371K LUTs, compared to the 166K LUTs needed for the finite-window version.
However, due to the vertex-level parallelism, the decoding latency has a modest increase from 194 ns to 230 ns.
Thus, even with sliding window decoding, the decoder is 56 times faster than the rate of measurement.

\subsection{Quantitative comparison with related work}
\label{ssec:comparison}

Our empirical results as shown in \autoref{fig:d_variation} suggest that \name has a lower asymptotic complexity than any existing MWPM or UF implementation for which asymptotic complexities are available, e.g., ~\cite{fowler2014minimum,delfosse2017almost}. 
Indeed, the empirical results suggest that our decoder has a sub-linear time complexity: the decoding time per round decreases with the number of measurement rounds, which has never been achieved before.
This implies that \name can support arbitrarily large $d$ as the rate of decoding will always be faster than the rate of measurement.


\input{TABLE2}


In ~\autoref{tab:decoders} we compare our decoder with other hardware decoders in the literature that provide implementation-based results. We report the average decoding time for $d=13$ for decoders capable of decoding $d>13$. For other decoders, we report the average decoding time for the maximum $d$ it can support.

As seen in ~\autoref{tab:decoders},  the most notable prior implementation is the Collision Clustering decoder by Riverlane~\cite{barber2023realtime}. 
This decoder, like our work, is an alternative implementation of the Union-Find (UF) algorithm. 
Its novel approach involves each vertex tracking its growth and using a hardware-implemented function for efficient distance computation between vertices to decide which vertices should merge.
This reduces the memory access requirements for UF implementation, compared to similar prior designs like AFS~\cite{das2022afs}. 
The reuse of the distance calculation function in all merging operations results in substantially lower resource consumption for the Collision Clustering decoder compared to \name. 
For example, for $d=13$, the Collision Clustering decoder requires about 6K LUTs, whereas \name requires 340K LUTs.

However, the average case latency per measurement round in the Collision Clustering decoder increases linearly with $d$, creating an upper bound of maximum $d$ that can be decoded in real time. 
In contrast, \name can decode arbitrarily large $d$ as shown in this work. 
Furthermore, the speed of the decoder relies on the efficient calculation of the distance between any two nodes. 
While this is straightforward for unweighted edges, when weighted edges or erasure errors are present, the decoder requires a complex function such as Dijkstra's algorithm to calculate distances. 
Using such a function for distance calculation can significantly increase the decoding time, making decoding erasure errors or weighted edges faster than the rate of measurement likely prohibitive.

LILLIPUT~\cite{das2021liliput}, Astrea-G~\cite{Vittal2023Astrea} and WIT-Greedy~\cite{Liao2023WitGreedy} are not scalable for large $d$, due to their excessively high storage requirements.  
LILLIPUT \cite{das2021liliput} is a look-up-table-based decoder.  Look-up-table-based decoders can achieve fast decoding but are not scalable beyond $d=5$ as the look-up table size grows $O(2^{d^3})$. For $d=7$ surface code with seven measurement rounds, it would require a memory of $2^{168}$ Bytes, which is infeasible in any foreseeable future.
Astrea-G~\cite{Vittal2023Astrea} and WIT-Greedy~\cite{Liao2023WitGreedy} store weights of all pairs of vertices and compare probable matchings. Astrea-G uses a greedy algorithm to pre-select matchings, and WIT-Greedy selects the least weight matching directly using a greedy algorithm, reducing accuracy further. 
The memory requirement for their weight tables grows $O(d^6)$, limiting their implementations at $d=9$ and $d=11$, respectively.
In contrast, our work has successfully demonstrated the implementation of a $d=51$ surface code on a VCU129 FPGA.
Furthermore, while these decoders could be adapted to circuit-level noise, accommodating erasure errors would exacerbate their already substantial memory requirements, owing to the need to process additional erasure inputs.

Overwater et al.~\cite{overwater2022NN} implements a neural network-based decoder. 
As shown by the authors, the decoder requires 44K LUTs for distance 5 for a single measurement round.
This worsens with distance as the input layer scales $O(d^2)$ with distance and $O(d^3)$ if $d$ rounds of measurements are considered. 
In comparison, \name with $d=5$ with 5 measurement rounds requires only 11K LUTs.

Our decoder outperforms the two fastest software MWPM decoder, Sparse Blossom~\cite{higgott2023sparse} and Fusion Blossom~\cite{wu2023Fusion}, by an order of magnitude. 
According to our evaluation, Sparse Blossom and Fusion Blossom take 160 ns and 295 ns per measurement round, respectively, for $d=13$ under $p=0.1\%$ phenomenological noise, using a single core of an M1 Max processor. 
In contrast, \name achieves an average decoding time of 15 ns per measurement round under the same conditions, which is more than 60 times faster than the current state-of-the-art measurement rate~\cite{Chen2021Exponential}.

\balance

%% file: TABLE2.tex
\begin{table*}
    \centering
    \begin{tabular}{|l|c|r|r|r|r|c|c|c|}
         \hline
         \multirow{4}{*}{Name} & \multirow{4}{*}{Algorithm} & \multicolumn{2}{c|}{Max.} & \multicolumn{2}{c|}{Decoding Time (ns)} & \multicolumn{3}{c|}{Features} \\
         \cline{3-9}
         & & \multirow{3}{*}{$d$} & \multirow{3}{*}{$m$} & \multirow{3}{*}{Avg.} & \multirow{3}{*}{Per Round} & \multicolumn{3}{c|}{\small{(\checkmark = implemented, $\ast$ = easily extendable)}} \\
         \cline{7-9}
         & & & & & & Erasures & Weighted & Circuit \\
         & & & & & & & edges & noise \\
         \hline\hline
         LILLIPUT~\cite{das2021liliput} & Look-up table & 5 & 2 & 42 & 21.0 & & \checkmark & $\ast$ \\ \hline
         Overwater et al.~\cite{overwater2022NN} & Neural Network & 5 & 1 & 88 & 87.6 & $\ast$ & \checkmark & $\ast$ \\ \hline
         WIT-Greedy~\cite{Liao2023WitGreedy} & Greedy & 11 & $d$ & 370 & 33.6 & & \checkmark & $\ast$ \\ \hline
         Astrea-G~\cite{Vittal2023Astrea} & Greedy & 9 & $d$ & 450 & 50.0 & & \checkmark & $\ast$ \\ \hline
         Collision Clustering~\cite{barber2023realtime} & Union-Find & 21 & $d$ & 2000 & 160.0 & & & \checkmark \\ \hline 
         \hline
         \textbf{Helios} (This work) & Union-Find & 51 & $d$ & 250 & 19.3 & \checkmark & \checkmark & \checkmark \\ \hline
    \end{tabular}
    \caption{Implementations of Surface Code Decoders on Classical Hardware (FPGA): Average decoding time is provided for $d=13$ for decoders capable of decoding $d \geq 13$ and for maximum $d$ for other decoders. Max $d$ is the maximum $d$ the decoder can decode faster than the rate of measurement within the device resource budget. Max $m$ is the maximum number of measurement rounds considered for decoding. }
    \label{tab:decoders}
\end{table*}

%% file: related_work.tex
\section{Related Work}
\label{sec:related}

There is a large body of literature on fast QEC decoding, e.g., \cite{battistel2023realtime, Terhal2015quantum, Gottesman2009Introduction, 2013LidarQuantumTopological}. The most related are solutions that leverage parallel computing resources.

Fowler~\cite{fowler2014minimum} describes a method for decoding at the rate of measurement ($O(d)$).
The proposed design divides the decoding graph among specialized hardware units arranged in a grid. 
Each unit contains a subset of vertices and can independently decode error chains contained within it. 
The design is based on the observation that large error patterns spanning multiple units are exponentially rare, so inter-unit communication is not frequently required.  
It, however, paradoxically assumes that the number of vertices per unit is ``sufficiently large,'' and a unit can find an MWPM for its vertices within half the measurement time on average. Not surprisingly, to date, no implementation or empirical data have been reported for this work.
Our approach uses vertex-level parallelism and leverages the same observation that communication between distant vertices is infrequent.

NISQ+\cite{Holmes2020Nisq} and QECOOL\cite{Ueno2021Qecool} parallelize computation at the ancilla level, where a single compute unit handles all vertices in the decoding graph representing measurements of one ancilla. 
This results in an increase in decoding time per measurement round as $d$ increases.
In contrast, we allocate a processing element per vertex, which results in decreasing decoding time per measurement round with $d$ at the expense of the number of parallel units growing $O(d^3)$.
Furthermore, they both implement the same greedy decoding algorithm, which is much lower in accuracy than the UF decoder used in this work.
QECOOL has an accuracy of approximately four orders of magnitude lower than a UF decoder~\cite{das2022afs}, and NISQ+ ignores measurement errors, further lowering its accuracy than QECOOL.

Skoric et al.~\cite{Skoric2022Parallel}, Tan et al.~\cite{tan2022scalable} and Wu~\cite{wu2023Fusion} propose similar methods of using measurement round-level parallelism, in which a decoder waits for a large number of measurement rounds to be completed and then decodes multiple blocks of measurement rounds in parallel. 
By using sufficient parallel resources, these methods can achieve a faster decoding rate than the measurement rate.
However, the latency of such approaches grows with the number of measurement rounds the decoder needs to batch to achieve a throughput equal to the rate of measurement.
In contrast, our approach exploits vertex-level parallelism and completes the decoding of every $d$ round of measurements with an average latency that grows sublinearly with $d$.

Since the initial release of this work~\cite{liyanage2023scalable}, two alternative designs employing vertex-level parallelism have been reported:  Actis~\cite{Chan2023Actis} and Heer et al.~\cite{Heer2023NovelUD}. 
Both map each vertex to a processing element and support nearest-neighbor communication. 
However, unlike our approach, these designs incorporate communication of processing elements with the central controller through the vertex array, resulting in a notable increase in coordinating overhead.
Furthermore, no implementation has been reported for either of them, making a direct comparison in terms of resource usage difficult.

Pipelining can be considered a special form of using compute resources in parallel, i.e., in different pipeline stages. 
Examples include AFS \cite{das2022afs},  LILLIPUT \cite{das2021liliput}, Astrea-G \cite{Vittal2023Astrea} and Collision Clustering \cite{barber2023realtime}.
However, pipelining is limited in how much parallelism it can leverage: the number of pipeline stages.
This results in a maximum $d$, which they can decode faster than the rate of measurement. 
The largest $d$ reported for pipelined decoders is $d=23$, which the ASIC design described in \cite{barber2023realtime} achieves with $240 ns$ per measurement round.
The parallelism of our decoder grows along $d^3$, which enables us to achieve a sublinear average case latency, including decoding $d=23$ within $24.1 ns$.
However, \name uses significantly more resources due to increased parallelism.

%% file: conclusion.tex
\section{Conclusion}
\label{sec:Conclusion}

We describe a distributed design for the Union Find decoder for quantum error-correcting surface codes, along with \name, a system architecture for its realization. 
Our FPGA-based implementation of \name demonstrates empirically that the average decoding time grows sub-linearly with the $d$. Using a VCU129 FPGA, \name decodes distance 21 surface codes at an average speed of 11.5 ns per measurement round, the fastest to the best of our knowledge.
\name is faster and more scalable than any previously reported surface code decoder implementations. 
Furthermore, to address resource constraints, \name can efficiently reuse FPGA resources, albeit with increased latency. 
We experimentally demonstrate that \name can decode extremely large surface codes such as $d=51$ on a VCU 129 FPGA which validates that \name can support surface code of any useful distance.


%% file: fpga_algo.tex
\section{FPGA-oriented algorithm}

\IncMargin{0.4em}
\begin{algorithm}[!h]
\setcounter{AlgoLine}{95}
\DontPrintSemicolon
\caption{FPGA-oriented algorithm for vertex $v$ in the distributed UF decoder.}\label{alg:FPGA_distributed_uf}
\footnotesize
$v.cid \gets v.id$;
$v.odd  \gets v.m$;
$v.parent \gets v.id$;
$v.st\_odd \gets v.m$ \;
~\\
\% Stage transition logic\\
\Posedge{}{
\lIf{$\code{global\_stage} = $\emph{\code{terminate}}}{
 \Return
}
\lElseIf{$\code{global\_stage} = $\emph{\code{growing}}}{
    $v.\code{stage} \gets \code{growing}$ \label{line:PE_move_growing} 
}
\lElseIf{$v.\code{stage} = $\emph{\code{growing}}}{
    $v.\code{stage} \gets \code{merging}$
    \label{line:PE_move_merging} 
}
}
~\\
\% Growing logic\\
\Posedge{}{
    \If{$v.\code{stage} = $\emph{\code{growing}}}{
         \ForEach{\textup{$e=\langle u,v\rangle\in v.E$ and $v.id < u.id$}}{
                \If{$e.$\emph{\code{growth}}$<e.w$ \textup{and} $u.cid\neq v.cid$}{ \label{line:FPGA_comparenupdate} 
                    \If{v.odd \textup{and} u.odd}{
                        $e.$\code{growth}$\gets \text{MIN}(e.$\code{growth}$+2$, $w$)
                    }
                    \ElseIf{v.odd \textup{or} u.odd}{
                        $e.$\code{growth}$\gets \text{MIN}(e.$\code{growth}$+1$, $w$)
                    }
                }
        }
    }
}
~\\
\% Merging logic \label{line:FPGA_merging}\\
\Posedge{}{
    Let $u$ be $\arg\min_{u \in (v.nb~\cup~\{v\})}(u.cid)$\\
    \If{$u.cid<v.cid$}{ 
        $v.cid \gets u.cid$  \\
        $v.parent \gets u.id$ 
    }
}
~\\
\Posedge{}{
$v.st\_odd \gets subtree\_parity(v)$
}
~\\
 \Posedge{}{       
    \lIf{$v.parent = v.id$}{
            $v.odd \gets v.st\_odd$ 
    }
    \lElse{
        \textup{$v.odd \gets u.odd$ where $u.id = v.parent$} 
    }
}
~\\
\% Checking logic \label{line:FPGA_checking}\\
\Posedge{}{
    \If{\textup{$ \exists u \in v.\code{nb}, (u.cid \neq v.cid\ \|\ v.odd \neq u.odd)$}}{
        $v.\code{busy} \gets \code{true}$\\
    }
    \ElseIf{\textup{ $v.st\_odd \neq subtree\_parity(v)$}}{
        $v.\code{busy} \gets \code{true}$\\
    }
    \ElseIf{\textup{ $(v.parent =  v.id$ \& $v.odd \neq v.st\_odd )$}}{
        $v.\code{busy} \gets \code{true}$\\
    }
    \Else{$v.\code{busy} \gets \code{false}$}
}
~\\
\function {subtree\_parity($v$)}{
$parity \gets v.m$\\
    \ForEach{$u \in v.\code{child}$}{
        $parity \gets \text{XOR}(parity, u.st\_odd)$
    }
\Return parity
}

\end{algorithm}

\begin{algorithm} [!h]
\DontPrintSemicolon 
\setcounter{AlgoLine}{161}
\footnotesize
\caption{FPGA-oriented controller logic}\label{alg:FPGA_gc}
$\code{global\_stage} \gets \code{growing}$ \\
\Posedge{}{
    \If{\textup{$\code{global\_stage} = \code{growing}$}} {
        $\code{global\_stage} \gets \code{merging}$ \\
        \%Wait until all PEs are in Merging Stage \\
        Wait 2 clock cycles
    }
    \ElseIf{ \textup{$\forall v \in V, v.\code{busy} = \code{false}$}} {
        \If{ \textup{$\forall v \in V, v.codd = \code{false}$}} {
            $\code{global\_stage} \gets \code{terminate}$ \label{line:FPGA_terminate}\\
        }
        \Else{
            $\code{global\_stage} \gets \code{growing}$
        }
    }
}
\end{algorithm}
\DecMargin{0.4em}

In ~\autoref{alg:FPGA_distributed_uf} and ~\autoref{alg:FPGA_gc} we show the FPGA-oriented algorithm for distributed UF.